\documentclass[runningheads,openright]{svmult}

\usepackage{makeidx}   
\usepackage{graphicx}  
\usepackage{subeqnar}  
\usepackage{multicol}  
\usepackage{physprbb}  
\makeindex             
\input psfig.sty

\newcommand{\age}{\mathrel{\hbox{\rlap{\hbox{\lower4pt\hbox{$\sim$}}}\hbox{$>$}}}}
\newcommand{\cmq}{\mbox{~cm$^{-3}$}}

\newcommand{\eg}{e.g., } 
\newcommand{\ergsec}{\mbox{~erg s$^{-1}$}}

\newcommand{\ergcm}{\mbox{~erg cm$^{-2}$}}

\newcommand{\etal}{et al.}

\newcommand{\gae}{\mathrel{>\kern-1.0em\lower0.9ex \hbox{$\sim$}}}

\newcommand{\gsim}{\!\!\!\phantom{\ge}\smash{\buildrel{}\over {\lower2.5dd\hbox{$\buildrel{\lower2dd\hbox{$\displaystyle>$}}\over \sim$}}}\,\,}
\newcommand{\gtap}{\mathrel{\hbox{\rlap{\lower.55ex \hbox {$\sim$}}\kern-.3em \raise.4ex \hbox{$>$}}}}
\newcommand{\gtrsim}{\mathrel{\hbox{\rlap{\hbox{\lower4pt\hbox{$\sim$}}}\hbox{$>$}}}}

\newcommand{\ie}{i.e., }
\newcommand{\keV}{\mbox{~keV}}
\newcommand{\kmsMpc}{\mbox{~km s$^{-1}$ Mpc$^{-1}$}}

\newcommand{\lae}{\mathrel{<\kern-1.0em\lower0.9ex \hbox{$\sim$}}}
\newcommand{\lesssim}{\mathrel{\hbox{\rlap{\hbox{\lower4pt\hbox{$\sim$}}}\hbox{$<$}}}}
\newcommand{\lsim}{\!\!\!\phantom{\le}\smash{\buildrel{}\over {\lower2.5dd\hbox{$\buildrel{\lower2dd\hbox{$\displaystyle<$}}\over \sim$}}}\,\,}

\newcommand{\ltap}{\mathrel{\hbox{\rlap{\lower.55ex \hbox {$\sim$}} \kern-.3em \raise.4ex \hbox{$<$}}}}
\newcommand{\ltsima}{$\; \buildrel < \over \sim \;$}

\newcommand{\Msunsec}{\mbox{~M$_\odot$ s$^{-1}$}}

\newcommand{\Msun}{\mbox{~M$_\odot$}}

\newcommand{\simlt}{\lower.5ex\hbox{\ltsima}} 

\newcommand{\yr}{\mbox{~yr}} 
 
\def\AGASA{{\sl Akeno Giant Air Shower Array (AGASA)}\index{instruments, agencies, observatories, and programs!Akeno Giant Air Shower Array (AGASA)}}

\def\ASCA{{\sl Advanced Satellite for Cosmology and Astrophysics (ASCA)}\index{instruments, agencies, observatories, and programs!Advanced Satellite for Cosmology and Astrophysics (ASCA)}}

\def\B{{\sl BeppoSAX}\index{instruments, agencies, observatories, and programs!BeppoSAX}}
\def\BATSE{{\sl Burst and Transient Source Experiment (BATSE)}\index{instruments, agencies, observatories, and programs!Compton Gamma-Ray Observatory (CGRO)!Burst and Transient Source Experiment (BATSE)}}
\def\BA{{\sl BATSE}\index{instruments, agencies, observatories, and programs!Compton Gamma-Ray Observatory (CGRO)!Burst and Transient Source Experiment (BATSE)}}

\def\E{{\sl Einstein}\index{instruments, agencies, observatories, and programs!Einstein Observatory -- HEAO-2 (Einstein)}}

\def\HIRES{{\sl High Resolution Fly's Eye (HiRes)}}
\def\HiRes{{\sl HiRes}}

\def\I{{\sl IRAS}}

\def\WFPC2{{\sl Wide-Field Planetary Camera 2 (WFPC2)}\index{instruments, agencies, observatories, and programs!Hubble Space Telescope (HST)!Wide-Field Planetary Camera 2 (WFPC2)}}
\def\WFPc2{{\sl WFPC2}\index{instruments, agencies, observatories, and programs!Hubble Space Telescope (HST)!Wide-Field Planetary Camera 2 (WFPC2)}}

\def\YAK{{\sl Yakutsk Extensive Air Shower Array (Yakutsk)}}

\def\ISM{{interstellar medium (ISM)}\index{interstellar medium}}
\def\ISm{{ISM}\index{interstellar medium}}

\def\UHECR{{ultra-high energy cosmic-rays (UHECR)}\index{cosmic-rays!ultra-high energy}}
\def\UHE{{UHECR}\index{cosmic-rays!ultra-high energy}}

\def \aap #1 #2 {{Astron. Astrophys.\/} {\bf #1}, #2~}
\def \aar #1 #2 {{Astron. Astrophys. Rev.\/} {\bf #1}, #2~}
\def \aas #1 #2 {{Astron. Astrophys. Suppl. Ser.\/} {\bf #1}, #2~}
\def \aj #1 #2 {{Astron. J.\/} {\bf #1}, #2~}
\def \al #1 #2 {{Astron. Lett.\/} {\bf #1}, #2~}
\def \an #1 #2 {{Astron. Nach.\/} {\bf #1}, #2~}
\def \annap #1 #2 {{Annals Ap.\/} {\bf #1}, #2~}
\def \aph #1 {{astro-ph\/} {#1}~}
\def \ar #1 #2 {{Astron. Rep.\/} {\bf #1}, #2~}
\def \araap #1 #2 {{Ann. Rev. Astron. Astrophys.\/} {\bf #1}, #2~}
\def \asiagoc #1 #2 {{Asiago Contr.\/} {\bf #1}, #2~}
\def \apj #1 #2 {{Astrophys. J.\/} {\bf #1}, #2~}
\def \apjl #1 #2 {{Astrophys. J. Lett.\/} {\bf #1}, #2~}
\def \apjs #1 #2 {{Astrophys. J. Suppl.\/} {\bf #1}, #2~}
\def \apjsub #1 {{Astrophys. J.\/} {#1}~}
\def \apph #1 #2 {{Astropart. Phys.\/} {\bf #1}, #2~}
\def \apss #1 #2 {{Astrophys. Space Sci.\/} {\bf #1}, #2~}
\def \aspc #1 #2 {{ASP Conf.~Proc.\/} {\bf #1}, #2~}
\def \aspl #1 {{ASP Leaflet\/} {#1}~}
\def \asr #1 #2 {{Adv. Space Res.\/} {\bf #1}, #2~}
\def \astrl #1 #2 {{Astron. Lett.\/} {\bf #1}, #2~}
\def \azh #1 #2 {{Astron. Zhurnal\/} {\bf #1}, #2~}
\def \baas #1 #2 {{Bull. Am. Astron. Soc.\/} {\bf #1}, #2~}
\def \ban #1 #2 {{Bull. Astron. Inst. Neth.\/} {\bf #1}, #2~}
\def \basi #1 #2 {{Bull. Astron. Soc. India\/} {\bf #1}, #2~}
\def \ca #1 #2 {{Chinese Astron.\/} {\bf #1}, #2~}
\def \coap #1 #2 {{Contrib. Oss. Astrofis. Padova in Asiago\/} {\bf #1}, #2~}
\def \cap #1 #2 {{Comm. Astrophys.\/} {\bf #1}, #2~}
\def \emsg #1 {{ESO Messenger\/} {#1}~}
\def \gcn #1 {{GCN\/} {#1}~}
\def \hast #1 #2 {{Highlights of Astronomy\/} {\bf #1}, #2~}
\def \iauc #1 {{IAUC\/} {#1}~}
\def \iaus #1 #2 {{IAU Symp. 110: VLBI \& Compact Radio Sources\/} {\bf #1}, #2~}

\def \jcam #1 #2 {{J. Comp. Appl. Math.\/} {\bf #1}, #2~}
\def \jet #1 #2 {{JETP Lett.\/} {\bf #1}, #2~}
\def \jha #1 #2 {{J. Hist. Astron.\/} {\bf #1}, #2~}
\def \jrasc #1 #2 {{J. R. Astron. Soc. Canada\/} {\bf #1}, #2~}
\def \mem #1 #2 {{Mem. R. Astron. Soc.\/} {\bf #1}, #2~}
\def \mess #1 #2 {{The Messenger\/} {\bf #1}, #2~}
\def \mnras #1 #2 {{Mon. Not. R. Astron. Soc.\/} {\bf #1}, #2~}
\def \mplb #1 #2 {{Mod. Phys. Lett. B\/} {\bf #1}, #2~}
\def \nat #1 #2 {{Nature\/} {\bf #1}, #2~}
\def \newa #1 #2 {{New Astron.\/} {\bf #1}, #2~}
\def \nuca #1 #2 {{Nucl. Phys. A\/} {\bf #1}, #2~}
\def \nucb #1 #2 {{Nucl. Phys. B\/} {\bf #1}, #2~}
\def \npps #1 #2 {{Nucl. Phys. Proc. Suppl.\/} {\bf #1}, #2~}
\def \nyasa #1 #2 {{NY Acad. Sci. Ann.\/} {\bf #1}, #2~}
\def \obsy #1 #2 {{The Observatory\/} {\bf #1}, #2~}
\def \phfl #1 #2 {{Phys. Fluids\/} {\bf #1}, #2~}
\def \phytd #1 #2 {{Phys. Today\/} {\bf #1}, #2~}
\def \prl #1 #2 {{Phys. Rev. Lett.\/} {\bf #1}, #2~}
\def \prp #1 #2 {{Phys. Rep.\/} {\bf #1}, #2~}
\def \phyr #1 #2 {{Phys. Rev.\/} {\bf #1}, #2~}
\def \phyrd #1 #2 {{Phys. Rev. D\/} {\bf #1}, #2~}
\def \prasa #1 #2 {{Proc. Astron. Soc. Australia\/} {\bf #1}, #2~}
\def \pasa #1 #2 {{Pub. Astron. Soc. Australia\/} {\bf #1}, #2~}
\def \pasj #1 #2 {{Pub. Astron. Soc. Japan\/} {\bf #1}, #2~}
\def \pasp #1 #2 {{Pub. Astron. Soc. Pacific\/} {\bf #1}, #2~}
\def \qjras #1 #2 {{Q. J. R. Astron. Soc.\/} {\bf #1}, #2~}
\def \rma #1 #2 {{Rev. Mod. Astron.\/} {\bf #1}, #2~}
\def \rpp #1 #2 {{Rep. Prog. Phys.\/} {\bf #1}, #2~}
\def \rpph #1 #2 {{Rev. Plasma Phys.\/} {\bf #1}, #2~}
\def \sait #1 #2 {{Mem.\ Soc.\ Astron.\ It.\/} {\bf #1}, #2~}
\def \sast #1 #2 {{Sov. Astron.\/} {\bf #1}, #2~}
\def \sal #1 #2 {{Sov. Astron. Lett.\/} {\bf #1}, #2~}
\def \sat #1 #2 {{Sky \& Tel.\/} {\bf #1}, #2~}
\def \sci #1 #2 {{Science\/} {\bf #1}, #2~}
\def \spie #1 #2 {{SPIE\/} {\bf #1}, #2~}
\def \shns #1 #2 {{Stud. Hist. Nat. Sci.\/} {\bf #1}, #2~}
\def \va #1 #2 {{Vist. Astron.\/} {\bf #1}, #2~}
\newcommand{\GRB}[1]{GRB#1\index{GRB!individual!GRB#1}}

\newcommand{\SN}[1]{SN#1\index{supernova!individual!SN#1}}

\index{agency|see {instruments, agencies, observatories, and programs}}
\index{Baade-Wesselink method|see {supernova, distance, Baade-Wesselink method}}
\index{blastwave|see {(supernova, shock, circumstellar) or (GRB, shock, external)}}
\index{circumstellar shock|see {supernova, shock, circumstellar}}
\index{CLUMPYSYN|see {supernova, model, CLUMPYSYN}}
\index{collapsar|see {GRB, progenitor, collapsar}}
\index{dark burst|see {GRB, optical, dark}}
\index{expanding photosphere method|see {supernova, distance, expanding photosphere method}}
\index{filled-center supernova remnant|see {supernova remnant, plerionic}}
\index{forward shock|see {(supernova, shock, circumstellar) or (GRB, shock, external)}}
\index{GRB!model!interstellar medium interactor|see {GRB, circumburst medium, interstellar medium}}
\index{GRB!model!wind interactor|see {GRB, circumburst medium, stellar wind}}
\index{guest star|see {supernova, historical}}
\index{GRB!progenitor!hypernova|see {hypernova}}
\index{GRB!progenitor!Wolf-Rayet star|see {star, Wolf-Rayet}}
\index{interstellar scintillation|see {GRB, radio, scintillation}}
\index{ML MONTE CARLO|see {supernova, model, ML MONTE CARLO}}
\index{NLTE (non-local thermodynamic equilibrium)|see {supernova, model, NLTE}}
\index{outgoing shock|see {(supernova, shock, circumstellar) or (GRB, shock, external)}}
\index{observatory|see {instruments, agencies, observatories, and programs}}
\index{PHOENIX|see {supernova, model, PHOENIX}}
\index{presupernova wind|see {supernova, progenitor, wind}}
\index{program|see {instruments, agencies, observatories, and programs}}
\index{pulsar|see {star, neutron}}
\index{radio supernova|see {supernova, radio}}
\index{RL NEBULAR|see {supernova, model, RL NEBULAR}}
\index{Sobolev approximation|see {supernova, model, Sobolev approximation}}
\index{supernova!cosmology|see {cosmology}}
\index{synchrotron self-absorption|see {supernova, synchrotron, self-absorption}}
\index{SYNOW|see {supernova, model, SYNOW}}
\index{synthetic spectrum|see {supernova, model, synthetic spectra}}
\index{VLBI|see {instruments, agencies, observatories, and programs, VLBI}}

\begin{document}


\mainmatter
\title*{Gamma-Ray Bursts: The Underlying Model
}
\toctitle{Gamma-Ray Bursts: The Underlying Model}
%
%
\titlerunning{Gamma-Ray Bursts: The Underlying Model}
%
\author{Eli Waxman}
\authorrunning{Waxman}
%
%
\institute{Weizmann Institute of Science, Rehovot 76100, Israel}

\maketitle              

\begin{abstract}
A pedagogical derivation is presented of the ``fireball'' model of 
$\gamma$-ray bursts, according to which the observable effects 
are due to the dissipation of the kinetic energy
of a relativistically expanding wind, a ``fireball.'' The main open 
questions are emphasized, and key afterglow observations, that provide 
support for this model, are briefly discussed. The relativistic outflow is, 
most likely, driven by the 
accretion of a fraction of a solar mass onto a newly born (few) solar mass 
black hole. The observed radiation is produced
once the plasma has expanded to a scale much larger than that of the
underlying ``engine,'' and is therefore largely independent of the details
of the progenitor, whose gravitational collapse leads to fireball formation. 
Several progenitor scenarios, and the prospects for discrimination among
them using future observations, are discussed. 
The production in $\gamma$-ray burst fireballs of high energy protons and neutrinos, and the 
implications of burst neutrino detection by kilometer-scale telescopes under
construction, are briefly discussed.
\end{abstract}


\section{Introduction}

The widely accepted interpretation of the
phenomenology of $\gamma$-ray bursts (GRBs),
of $0.1-1$ MeV photons lasting for a few seconds
(see \cite{Fishman} for a review), is that the
observable effects are due to the dissipation of the kinetic energy
of a relativistically expanding wind, a 
``fireball\index{GRB!fireball},'' whose primal cause is not yet known. 
The recent detection of ``afterglows\index{GRB!afterglow},'' 
delayed low energy (X-ray\index{GRB!X-ray!emission} to radio\index{GRB!radio!emission}) emission 
of GRBs (see \cite{AG_ex_review} for a review),
confirmed the cosmological origin of the bursts
through the redshift determination of several GRB host-galaxies,
and supported standard model predictions of afterglows
that result from the collision of an expanding fireball with
its surrounding medium\index{GRB!circumburst medium} (see 
\cite{AG_th_review,fireballs2} for reviews). 

The fireball\index{GRB!fireball} model is described in 
\S\S\ \ref{sec:fireball-hyd}, \ref{sec:fireball-rad}, and \ref{sec:fireball-AG} of this chapter. 
The phenomenological arguments suggesting that
fireball formation is likely regardless of the underlying progenitor  
are presented, and fireball hydrodynamics and radiative processes are 
discussed in detail
in \S\S\ref{sec:fireball-hyd} and \ref{sec:fireball-rad}, respectively.
The main open questions related to fireball physics are discussed in
\S\ref{sec:openQs}. Since both the theory and the implications
of afterglow\index{GRB!afterglow} observations are
extensively discussed in other chapters of this volume, we include in 
\S\ref{sec:fireball-AG} of this chapter only a brief discussion of 
several key afterglow implications. We also limit the theoretical discussion
of fireball evolution\index{GRB!fireball!evolution} to the GRB production phase 
that precedes the afterglow phase during which evolution is
dominated by the interaction of the fireball with its surrounding medium\index{GRB!circumburst medium}.
We do discuss, however, the initial non-self-similar onset of this 
interaction, which marks the onset of the afterglow phase. 

The GRB progenitors\index{GRB!progenitor} are not yet known. We present in \S\ref{sec:progenitors}
the constraints imposed by observations on possible progenitors, and discuss
the (presently) leading candidates. Hints provided by afterglow observations,
which are extensively discussed in separate chapters of this volume, are
briefly reviewed.

The association of GRBs with \UHECR, the evidence for which is strengthened by recent 
afterglow observations, is based on two key points \cite{W95a}: 1) the constraints that a dissipative ultra-relativistic wind
must satisfy in order to allow acceleration of protons 
to energy $\sim10^{20}$~eV, the highest observed cosmic-ray energy,
are remarkably similar to the constraints imposed on the fireball wind 
by $\gamma$-ray observations, and 2) the
inferred local ($z = 0$) GRB energy generation rate of $\gamma$-rays
is remarkably similar to 
the local generation rate of \UHE\ implied by cosmic-ray observations.
We briefly discuss in \S\ref{sec:UHECR} production of high energy
protons\index{GRB!protons} and neutrinos\index{GRB!neutrinos} in GRB fireballs (see \cite{My-NPB-rev,My-LNP-rev}
for more detailed reviews). The GRB model for \UHE\ production makes unique predictions that may be
tested with operating and planned large area \UHE\ detectors 
\cite{HiRes,Auger1,TA,Auger2}\footnote{See also\emph{ http://www.physics.adelaide.edu.au/astrophysics/FlysEye.html}, \\ \emph{http://www-ta.icrr.u-tokyo.ac.jp/}, and \emph{http://www.auger.org/}}. In this review we focus, however,
on more recent predictions of neutrino emission, which may be tested
with planned high energy neutrino telescopes \cite{Halzen-rev99}. 
Detection of the predicted neutrino signal
will confirm the GRB fireball model for the production of \UHE\ and
may allow discrimination between different fireball progenitor scenarios.
Moreover, a detection of even a handful of neutrino events correlated 
with GRBs will allow a test for neutrino properties, \eg flavor
oscillation\index{GRB!neutrinos} and coupling to gravity\index{GRB!neutrinos}, with accuracy many orders of magnitude
better than currently possible.

\section{The Fireball Model: Hydrodynamics}
\label{sec:fireball-hyd}

\subsection{Relativistic Expansion}
\label{sec:Rel-expansion}

General phenomenological considerations, based on $\gamma$-ray
observations, indicate that,
regardless of the nature of the underlying sources, 
GRBs are produced by the dissipation of the kinetic energy of a 
relativistically expanding fireball\index{GRB!fireball}. 
The rapid rise time and short duration, $\sim1$~ms, 
observed in some bursts \cite{Bhat92,Fishman94} imply
that the sources are compact, with a linear scale comparable
to a light-ms, $r_0\sim10^7$~cm. 
The high $\gamma$-ray luminosity implied by cosmological distances, 
$L_\gamma\sim10^{52}$ \ergsec,
then results in a very high optical depth to pair creation since the energy of observed $\gamma$-ray photons is above
the threshold for pair production. The number
density of photons at the source $n_\gamma$ is approximately given by

\begin{equation}
L_\gamma = 4\pi r_0^2 cn_\gamma\epsilon,
\end{equation}
 
\noindent where $\epsilon\simeq1$MeV is the characteristic
photon energy. Using $r_0\sim10^7$cm, the optical depth for pair production
at the source is

\begin{equation}
\tau_{\gamma\gamma}\sim r_0 n_\gamma\sigma_T\sim{\sigma_TL_\gamma
\over4\pi r_0 c\epsilon}\sim10^{15},
\label{eq:tau-pair}
\end{equation}

\noindent where $\sigma_T$ is the Thomson cross section\index{GRB!fireball!Thomson cross section}.

The high optical depth implies that a thermal plasma 
of photons\index{GRB!fireball!photons}, electrons\index{GRB!fireball!electrons}, and positrons\index{GRB!fireball!positrons} is created, a fireball
which then expands and accelerates to 
relativistic velocities\index{GRB!fireball!relativistic} \cite{Goodman86-EW,Bohdan86}. 
The optical depth is reduced by relativistic expansion of the source.  If the source expands with a Lorentz factor\index{GRB!fireball!Lorentz factor} $\Gamma$,
the energy of photons in the source frame is smaller by a factor $\Gamma$
compared to that in the observer frame,
and most photons may therefore be below the pair production threshold.

A lower limit for $\Gamma$ may be obtained in the following way 
\cite{Baring,Krolik}. The GRB photon spectrum is well fitted in the \BATSE\ detectors 
range, 20~keV to 2~MeV \cite{Fishman}, by a combination
of two power-laws, $dn_\gamma/d\epsilon_\gamma\propto\epsilon_\gamma^{(\alpha-1)}$ ($\alpha$ is the flux density spectral index, $F_\nu \propto \nu^{+\alpha}$) 
with different values of $\alpha$ at low and high energy \cite{Band}.
Here, $dn_\gamma/d\epsilon_\gamma$ is the number of photons per unit 
photon energy. The
break energy (where $\alpha$ changes) in the observer frame is typically 
$\epsilon_{\gamma b}\sim1~{\rm MeV}$, 
with $\alpha \simeq 0$ at energies below the break and $\alpha \simeq -1$ 
above the break. In several cases, the spectrum has been observed to extend
to energies $>100$~MeV \cite{Fishman,GRB100MeV}.
Consider then a high energy test photon, with observed energy
$\epsilon_t$, trying to escape the relativistically expanding source.
Assuming that, in the source rest frame, the photon distribution is isotropic,
and that the spectrum of high energy photons follows 
$dn_\gamma/d\epsilon_\gamma\propto\epsilon_\gamma^{-2}$, 
the mean free path for pair
production (in the source rest frame) for a photon of energy 
$\epsilon'_{t} = \epsilon_t/\gamma$ (in the source rest frame) is

\begin{equation}
l_{\gamma\gamma}^{-1}(\epsilon'_{t})
 = {1\over2}{3\over16}\sigma_T\int d\cos\theta(1-\cos\theta)
\int_{\epsilon_{\rm th}(\epsilon'_{t},\theta)}^\infty d\epsilon
{U_\gamma\over2\epsilon^2}
 = {1\over16}\sigma_T{U_\gamma\epsilon'_{\rm t}\over(m_ec^2)^2} \,.
\label{eq:lgg}
\end{equation}

\noindent Here, $\epsilon_{\rm th}(\epsilon'_{t},\theta)$ is the minimum energy of
photons that may produce pairs\index{GRB!fireball!pair production} interacting with the test photon, 
given by $\epsilon_{\rm th}\epsilon'_t(1-\cos\theta)\ge2(m_ec^2)^2$ 
($\theta$ is the angle between the photons' momentum vectors).
$U_\gamma$ is the photon energy density\index{GRB!fireball!photons} (in the range
corresponding to the observed \BA\ range) in the source rest-frame,
given by 

\begin{equation}
L_\gamma = 4\pi r^2\gamma^2cU_\gamma.
\end{equation}

\noindent (Note that we have used a constant cross section, $3\sigma_T/16$, 
above the threshold $\epsilon_
{\rm th}$.) The cross section drops as $\log(\epsilon)/\epsilon$ for
$\epsilon\gg\epsilon_{\rm th}$; however, since the number density of
photons drops rapidly with energy, this does not introduce a large correction
to $l_{\gamma\gamma}$.

The source size constraint implied by the variability time is modified
for a relativistically expanding source\index{GRB!fireball!relativistic}. Since in the observer frame almost 
all photons propagate at a direction making an angle $<1/\Gamma$ with respect 
to the expansion direction, radiation seen by a distant observer
originates from a conical section of the source around the source-observer 
line of sight, with opening angle $\sim1/\Gamma$. Photons which are emitted
from the edge of the cone are delayed, compared to those emitted on the
line of sight, by $r/2\Gamma^2c$. Thus, the constraint on source size
implied by variability on time scale $\Delta t$ is 

\begin{equation}
r\sim2\Gamma^2c\Delta t. 
\label{eq:r_s}
\end{equation}

\noindent The time $r/c$ required for significant source expansion
corresponds to comoving time (measured in the source frame)
$t_{\rm co.}\approx r/\Gamma c$. The two-photon collision rate\index{GRB!fireball!two-photon collision rate} at the source
frame is $t_{\gamma\gamma}^{-1} = c/l_{\gamma\gamma}$. Thus, the source
optical depth to pair production\index{GRB!fireball!pair production} is 
$\tau_{\gamma\gamma} = t_{\rm co.}/t_{\gamma\gamma}\approx
r/\Gamma l_{\gamma\gamma}$. Using Eqs.~ (\ref{eq:lgg}) and (\ref{eq:r_s}) 
we have

\begin{equation}
\tau_{\gamma\gamma} = {1\over128\pi} {\sigma_TL_\gamma\epsilon_t
\over c^2(m_ec^2)^2\Gamma^6\Delta t}. 
\label{eq:taugg}
\end{equation}

\noindent Requiring $\tau_{\gamma\gamma}<1$ at $\epsilon_t$ 
we obtain a lower limit for $\Gamma$,

\begin{equation}
\Gamma\ge250\left[
L_{\gamma,52} \left({\epsilon_t
\over100{\rm MeV}}\right)\Delta t_{-2}^{-1}\right]^{1/6},
\label{eq:Gmingg}
\end{equation}

\noindent where $L_{\gamma} = 10^{52}L_{\gamma,52}$ \ergsec and 
$\Delta t = 10^{-2}\Delta t_{-2}$~s.

\subsection{Fireball Evolution}
\label{sec:fireball-evolution}

As the fireball\index{GRB!fireball!evolution} expands it cools, the photon
temperature $T_\gamma$ 
in the fireball frame decreases, and most pairs annihilate.
Once the pair density is sufficiently low, photons may escape.
However, if the observed radiation is due
to photons escaping the fireball as it becomes optically thin, two 
problems arise. First, the photon spectrum is quasi-thermal,
in conflict with observations. Second, 
the source size, $r_0 \sim 10^7$~cm, and the total energy emitted in 
$\gamma$-rays,
$\sim10^{53}$ erg, suggest that the underlying energy source is related
to the gravitational collapse of a $\sim1$ \Msun\ object.
Thus, the plasma is expected to be loaded
with baryons\index{GRB!fireball!baryons} which may be injected with the radiation or present in the 
atmosphere surrounding the source. A small baryonic load, $\geq10^{-8}$ \Msun, increases the optical depth due to Thomson scattering\index{GRB!fireball!Thomson scattering}
on electrons associated with the loading protons, so that 
most of the radiation energy is converted to
kinetic energy of the relativistically\index{GRB!fireball!relativistic} expanding baryons before the plasma
becomes optically thin \cite{kinetic1,kinetic2}. 
To overcome both problems it was proposed \cite{RnM92} that the
observed burst is produced once the kinetic energy of the ultra-relativistic 
ejecta\index{GRB!ejecta} is re-randomized by some dissipation process at large radius, beyond
the Thomson photosphere\index{GRB!fireball!Thomson photosphere}, and then radiated as $\gamma$-rays. Collision 
of the relativistic baryons
with the \ISM\index{GRB!circumburst medium}\ \cite{RnM92}, and 
internal collisions within the ejecta itself 
\cite{internal3,internal1,internal2}, 
were proposed as possible dissipation processes.
Most GRBs show variability on time scales much shorter (typically $10^{-2}$ times) than the total GRB duration. Such variability is hard to explain in models where the
energy dissipation is due to external shocks \cite{SnP_var,Woods95}.
Thus, it is believed that internal collisions are responsible for the
emission of $\gamma$-rays. 

Let us first consider the case where the energy release from the source
is ``instantaneous,'' \ie on a time scale of $r_0/c$. We assume that most of 
the energy is released in the form of photons, \ie that the fraction of 
energy carried by baryon rest mass $M$ satisfies
$\eta^{-1} \equiv Mc^2/E\ll1$. 
The initial thickness of the fireball shell
is $r_0$. Since the plasma accelerates to relativistic\index{GRB!fireball!relativistic}
velocity, all fluid elements move with velocity close to $c$, and
the shell thickness remains constant at $r_0$ (this changes at very late
time, as discussed below).
We are interested in the stage where the optical depth (due to pairs and/or
electrons associated with baryons) is high, but only a small fraction of
the energy is carried by pairs. 

The entropy of a fluid component with zero chemical potential is 
$S = V(e+p)/T$, where $e$, $p$ and $V$ are the (rest frame) energy density, 
pressure and volume. For the photons $p = e/3\propto T_\gamma^4$.
Since initially both the rest mass and thermal energy of
baryons\index{GRB!fireball!baryons} is negligible, the entropy is provided by the photons\index{GRB!fireball!photons}.
Conservation of entropy implies

\begin{equation}
r^2 \Gamma(r) r_0 T_\gamma^3(r) = {\rm const} ,
\label{eq:entropy}
\end{equation}

\noindent and conservation of energy implies

\begin{equation}
r^2 \Gamma(r) r_0 \Gamma(r) T_\gamma^4(r) = {\rm const}\,.
\label{eq:energy}
\end{equation}

\noindent Here $\Gamma(r)$ is the shell Lorentz factor. 
Combining (\ref{eq:entropy}) and (\ref{eq:energy}) we find

\begin{equation}
\Gamma(r) \propto r, \quad T_\gamma \propto r^{-1}, \quad n \propto r^{-3},
\label{eq:scaling1}
\end{equation}

\noindent where $n$
is the rest frame (comoving) baryon number density.

As the shell accelerates the baryon kinetic energy, $\Gamma Mc^2$,
increases. It becomes comparable to the total fireball energy when 
$\Gamma \sim \eta$, at radius $r_f = \eta r_0$.
At this radius most of the energy of the fireball is carried by the baryon
kinetic energy, and the shell does not accelerate further. 
Eq.~(\ref{eq:energy}) describing
energy conservation is replaced with 
$\Gamma = {\rm const}$. Eq.~(\ref{eq:entropy}), however, still holds. 
Eq.~(\ref{eq:entropy}) 
may be written as $T_\gamma^4/nT_\gamma = {\rm const}$ (constant entropy per 
baryon). This implies that the
ratio of radiation energy density to thermal energy density associated
with the baryons is $r$ independent.
Thus, the thermal energy associated with the baryons may be neglected at all 
times, and Eq.~(\ref{eq:entropy}) 
holds also for the stage where most of the fireball energy 
is carried by the baryon kinetic energy. Thus, for $r>r_f$ we have

\begin{equation}
\Gamma(r) = \Gamma\approx\eta,\quad T\propto r^{-2/3},\quad n\propto r^{-2}.
\label{eq:scaling2}
\end{equation}

Let us consider now the case of extended emission from the source, on time
scale $\gg r_0/c$. In this case, the source continuously emits energy
at a rate $L$, and the energy emission is accompanied by mass-loss rate 

\begin{equation}
\dot M = L/\eta c^2.
\end{equation}

\noindent For $r<r_f$ the fluid energy density is relativistic, $aT_\gamma^4/nm_pc^2 = \eta r_0/r$, and the speed of sound is $\sim c$. The time
it takes the shell at radius $r$ to expand significantly 
is $r/c$ in the observer frame, 
corresponding to $t_{\rm co.}\sim r/\Gamma c$ 
in the shell frame. During this time
sound waves can travel a distance 
$cr/\Gamma c$ in the shell frame, corresponding to $r/\Gamma^2 = r/(r/r_0)^2 = 
(r_0/r)r_0$ in the observer frame. This implies that at the early stages of
evolution, $r\sim r_0$, sound waves had enough time to smooth out spatial
fluctuations in the fireball over a scale $r_0$, but that regions separated by
$\Delta r>r_0$ cannot interact with each other. Thus, if the emission extends
over a time $t_{\rm GRB}\gg r_0/c$, 
a fireball of thickness $ct_{\rm GRB}\gg r_0$ would be formed,
which would expand as a collection of independent, roughly uniform, sub-shells
of thickness $r_0$. Each sub-shell would reach a final Lorentz factor\index{GRB!fireball!Lorentz factor} 
$\Gamma_f$, which may vary between sub-shells. This
implies that different sub-shells may have velocities differing by 
$\Delta v\sim c/2\eta^2$, where $\eta$ is some typical value representative
of the entire fireball. Different shells emitted at times differing
by $\Delta t$, $r_0/c<\Delta t<t_{\rm GRB}$, may therefore collide with
each other after a time $t_c\sim c\Delta t/\Delta v$, \ie at a radius

\begin{equation}
r_i\approx2\Gamma^2c\Delta t = 6\times10^{13}\Gamma^2_{2.5}\Delta t_{-2}
{\rm\ cm},
\label{eq:r_i}
\end{equation}

\noindent where $\Gamma = 10^{2.5}\Gamma_{2.5}$.
The minimum internal shock radius, $r\sim \Gamma^2 r_0$, 
is also the radius at which an individual sub-shell may experience
significant change in its width $r_0$, due to Lorentz factor variation
across the shell.

\subsection{The Allowed Range of Lorentz Factors and Baryon Loading}
\label{sec:eta}

The acceleration, $\Gamma \propto r$, of fireball plasma
is driven by radiation pressure. Fireball
protons\index{GRB!fireball!protons} are accelerated through their coupling to the electrons\index{GRB!fireball!electrons}, which
are coupled to fireball photons\index{GRB!fireball!photons}. 
We have assumed in the analysis presented above, that photons and electrons\index{GRB!fireball!electrons}
are coupled throughout the acceleration phase. However, if the baryon\index{GRB!fireball!baryons} loading
is too low, radiation may decouple from fireball electrons 
already at $r<r_f$. The fireball
Thomson optical depth\index{GRB!fireball!Thomson optical depth} is given by the product of comoving expansion time,
$r/\Gamma(r) c$, and the photon Thomson scattering rate\index{GRB!fireball!Thomson scattering}, $n_ec\sigma_T$.
The electron and proton comoving number densities are
equal, $n_e = n_p$, and are determined by equating the $r$ independent
mass flux carried by the wind, $4\pi r^2 c \Gamma(r) n_p m_p$, 
to the mass-loss rate\index{GRB!progenitor!mass-loss} from the underlying 
source, which is related to the rate $L$ at which energy 
is emitted through $\dot M = L/(\eta c^2)$. Thus,
during the acceleration phase, $\Gamma(r) = r/r_0$ and the 
Thomson optical depth, $\tau_T$, is $\tau_T\propto r^{-3}$.  The Thomson optical depth drops 
below unity at a radius $r<r_f = \eta r_0$ if $\eta>\eta_*$, where

\begin{equation}
\eta_* = \left({\sigma_T L \over 4\pi r_0 m_p c^3}\right)^{1/4} = 
       1.0\times10^3 L_{52}^{1/4}r_{0,7}^{-1/4}.
\label{eq:eta-star}
\end{equation}

\noindent Here $r_0 = 10^7r_{0,7}$~cm.

If $\eta>\eta_*$ radiation decouples from the fireball plasma at
$\Gamma = r/r_0 = \eta_*^{4/3}\eta^{-1/3}$. If $\eta\gg\eta_*$, then most of
the radiation energy is not converted to kinetic energy prior to radiation
decoupling, and most of the fireball energy escapes in the form of thermal
radiation. Thus, the baryon load of fireball shells, and the corresponding 
final Lorentz factors\index{GRB!fireball!Lorentz factor}, must be within the range 
$10^2\le\Gamma\approx\eta\le\eta_*\approx10^3$ 
in order to allow the production of the
observed non-thermal $\gamma$-ray spectrum. 

\subsection{Fireball Interaction with the Surrounding Medium}
\label{sec:fireball-interaction}

As the fireball expands, it drives a relativistic shock\index{GRB!shock!relativistic} (blastwave)
into the surrounding
gas, \eg into the \ISm\ gas if the explosion 
occurs within a galaxy. In what follows, we refer to the surrounding gas\index{GRB!circumburst medium}
as \ISm\ gas, although the gas need not necessarily be interstellar.
At early times, the fireball is little affected by the interaction with the 
\ISm. At late times, most of the fireball energy is transferred to the \ISm, and
the flow approaches the self-similar blastwave\index{GRB!fireball!self-similar solution} solution of Blandford and
McKee \cite{BnM76}. At this stage a single shock propagates into the \ISm,
behind which the gas expands with Lorentz factor\index{GRB!fireball!Lorentz factor}

\begin{equation}
\Gamma_{BM}(r) = \left(\frac{17E}{16\pi n m_p c^2}\right)^{1/2}r^{-3/2} = 
150\left({E_{53}\over n_0}\right)^{1/2}r_{17}^{-3/2},
\label{eq:Gamma_BM}
\end{equation}

\noindent where $E = 10^{53}E_{53}$ erg is the fireball energy,
$n = 1n_0{\rm\ cm}^{-3}$ is the \ISm\ number density, and $r = 10^{17}r_{17}$~cm
is the shell radius. 
The characteristic time
at which radiation emitted by shocked plasma at radius $r$
is observed by a distant observer is $t\approx r/4\Gamma_{BM}^2 c$
\cite{WAG-ring}.

The transition to self-similar expansion occurs on a time scale $t_{\rm SS}$ (measured
in the observer frame)
comparable to the longer of the two time scales set by the initial
conditions: the (observer) GRB duration $t_{\rm GRB}$ and the
(observer) time $t_{\Gamma}$ at which the self-similar Lorentz factor
equals the original ejecta Lorentz factor $\Gamma$,
$\Gamma_{BM}(t = t_\Gamma) = \Gamma$. Since $t = r/4\Gamma^2_{BM}c$,

\begin{equation}
t = \max\left[t_{\rm GRB}, 5\left({E_{53}\over
n_0}\right)^{1/3}\Gamma_{2.5}^{-8/3}{\rm\, s}\right].
\label{eq:t_tr}
\end{equation}

\noindent During the transition,  plasma shocked by the reverse shocks\index{GRB!shock!reverse}
expands with Lorentz factor\index{GRB!fireball!Lorentz factor} close to that given by the self-similar solution\index{GRB!fireball!self-similar solution},

\begin{equation}
\Gamma_{\rm tr.} \simeq \Gamma_{BM}(t = t_{\rm SS})
\simeq 245\left({E_{53}\over n_0}\right)^{1/8}t_1^{-3/8},
\label{eq:Gamma_tr}
\end{equation}

\noindent where $t = 10t_1$~s. The unshocked fireball
ejecta\index{GRB!ejecta!unshocked} propagate at the original expansion Lorentz factor, 
$\Gamma$,
and the Lorentz factor of plasma shocked by the reverse shock in
the rest frame of the unshocked ejecta is $\simeq\Gamma/\Gamma_{\rm tr.}$.
If $t \simeq t_{\rm GRB} \gg t_{\Gamma}$ then 
$\Gamma/\Gamma_{\rm tr.}\gg1$, the reverse shock is relativistic,
and the Lorentz factor associated with the random motion of protons 
in the reverse shock\index{GRB!shock!reverse} is $\Gamma_p^R\simeq\Gamma/\Gamma_{\rm tr.}$. 

If, on the other hand, $t \simeq t_{\Gamma}\gg t_{\rm GRB}$ then 
$\Gamma/\Gamma_{\rm tr.}\sim1$, and the reverse shock is not relativistic. 
Nevertheless, the following argument suggests that the reverse shock
speed is not far below $c$, and that the protons are therefore 
heated to 
relativistic energy, $\Gamma_p^R-1 \simeq 1$. The comoving time, measured
in the fireball ejecta\index{GRB!ejecta} frame prior to deceleration, 
is $t_{\rm co.} \simeq r/\Gamma c$. The expansion Lorentz
factor is expected to vary across the ejecta, $\Delta \Gamma/\Gamma \sim1 $,
due to variability of the underlying GRB source over the duration of 
its energy release.
Such variation would lead
to expansion of the ejecta, in the comoving frame, at relativistic speed\index{GRB!ejecta!relativistic}. 
Thus, at the deceleration radius, $t_{\rm co.} \simeq \Gamma t$,
the ejecta width exceeds $\simeq ct_{\rm co.} \simeq \Gamma c t$. Since
the reverse shock should cross the ejecta over a deceleration time 
scale, $\simeq\Gamma t$, the reverse shock speed must be close to 
$c$. We therefore conclude that 
the Lorentz factor associated with the random motion of protons 
in the reverse shock is approximately given by 
$\Gamma_p^R-1 \simeq \Gamma/\Gamma_{\rm tr.}$ for both 
$\Gamma/\Gamma_{\rm tr.}\sim1$
and $\Gamma/\Gamma_{\rm tr.}\gg1$. 

Since $t_{\rm GRB} \sim 10$~s is typically comparable to $t_\Gamma$, 
the reverse shocks are typically expected to be mildly relativistic.

\subsection{Fireball Geometry}
\label{sec:fireball-geometry}

We have assumed in the discussion so far that the fireball is spherically 
symmetric. However, a jet-like fireball behaves as if it were
a conical section of a spherical fireball as long as the jet\index{GRB!jet} opening
angle is larger than $\Gamma^{-1}$. This is due to the fact that
the linear size of causally connected regions, 
$ct_{\rm co.}\sim r/\Gamma$ in the fireball frame, corresponds to
an angular size $ct_{\rm co.}/r\sim\Gamma^{-1}$. Moreover, due
to the relativistic beaming\index{GRB!jet!relativistic} of radiation, a distant observer cannot
distinguish between a spherical fireball and a jet-like fireball, as
long as the jet opening angle $\theta>\Gamma^{-1}$. Thus, as long as 
we are discussing processes that occur when
the wind is ultra-relativistic, $\Gamma\sim300$ (prior to 
significant fireball deceleration by the surrounding medium), our
results apply for both a spherical and a jet-like fireball.
In the latter case, $L$ ($E$) in our
equations should be understood as the luminosity (energy) the fireball
would have carried had it been spherically symmetric.

\section{The Fireball Model: Radiative Processes}
\label{sec:fireball-rad}

\subsection{Gamma-Ray Emission}
\label{sec:fireball-gammas}

If the Lorentz factor\index{GRB!fireball!Lorentz factor} variability within the wind is significant,
internal shocks\index{GRB!shock!internal} will reconvert a substantial 
part of the kinetic energy to internal energy. The internal 
energy may then be radiated as 
$\gamma$-rays by synchrotron and inverse Compton emission\index{inverse Compton!emission} of
shock-accelerated electrons. The
internal shocks are expected to be mildly relativistic in the fireball 
rest frame, \ie characterized by Lorentz factor 
$\Gamma_i-1\sim$ a few. This is due to the fact that the allowed range
of shell Lorentz factors is $\sim10^2$ to $\sim10^3$ (see \S\ref{sec:eta}),
implying that the Lorentz factors associated with the relative velocities
are not very large. Since internal shocks are mildly
relativistic, we expect results related to particle
acceleration in sub-relativistic shocks (see \cite{Blandford87} for a review)
to be valid for acceleration
in internal shocks. In particular, electrons are 
expected to be accelerated to a power law energy distribution,
$dn_e/d\Gamma_e \propto \Gamma_e^{-p}$ for $\Gamma_e > \Gamma_m$, with $p \simeq 2$
\cite{AXL77,Bell78,BnO78}.

The minimum Lorentz factor $\Gamma_m$
is determined by the following consideration. 
Protons\index{GRB!fireball!protons} are heated in internal shocks
to random velocities (in the wind frame) 
$\Gamma_p^R-1 \approx \Gamma_i-1\approx1$. 
If electrons\index{GRB!fireball!electrons} carry a fraction $\xi_e$ of the shock internal energy, then
$\Gamma_m \approx \xi_e(m_p/m_e)$. The characteristic
frequency of synchrotron emission\index{GRB!synchrotron!emission} is determined by $\Gamma_m$ and 
by the strength of the magnetic field\index{GRB!magnetic field}. Assuming 
that a fraction $\xi_B$ of the internal energy 
is carried by the magnetic field\index{GRB!magnetic field},  
$4\pi r_i^2c\Gamma^2B^2/8\pi = \xi_B L_{\rm int.}$, 
the characteristic observed energy of synchrotron photons, 
$\epsilon_{\gamma b} = \Gamma\hbar\Gamma_m^2 eB/m_ec$, is

\begin{equation}
\epsilon_{\gamma b}\approx1\xi_B^{1/2}\xi_e^{3/2}{L_{\gamma,52}^{1/2}
\over\Gamma_{2.5}^2\Delta t_{-2}}~{\rm MeV}.
\label{eq:E_gamma}
\end{equation}

\noindent In deriving Eq.~(\ref{eq:E_gamma}) we have assumed that the
wind luminosity carried by internal plasma energy, $L_{\rm int.}$, is
related to the observed $\gamma$-ray luminosity through 
$L_{\rm int.} = L_\gamma/\xi_e$. This assumption is justified since
the electron synchrotron cooling\index{GRB!synchrotron!cooling} time is short
compared to the wind expansion time (unless the equipartition fraction 
$\xi_B$ is many orders of magnitude smaller than unity), and 
hence electrons lose all their energy radiatively. 
Fast electron cooling also results in a synchrotron
spectrum\index{GRB!synchrotron!spectrum} $dn_\gamma/d\epsilon_\gamma\propto\epsilon_\gamma^{-1-p/2} = 
\epsilon_\gamma^{-2}$ at $\epsilon_\gamma>\epsilon_{\gamma b}$,
consistent with observed GRB spectra \cite{Band}.

At present, there is no theory that allows the determination of 
the values of the equipartition\index{GRB!fireball!equipartition} fractions $\xi_e$ and $\xi_B$. 
Eq.~(\ref{eq:E_gamma}) implies that 
fractions not far below unity are required to account for the observed
$\gamma$-ray emission. We note that build up of magnetic field\index{GRB!magnetic field}
to near equipartition by electromagnetic instabilities is expected to
be a generic characteristic of collisionless shocks\index{GRB!shock!collisionless} 
(see the discussion in \cite{Blandford87} and references therein), and
is inferred to occur in other systems such as in supernova remnant shocks
(see, \eg \cite{Cargill88,Helfand87}).

\subsection{Break Energy Distribution}
\label{sec:fireball-break}

The $\gamma$-ray break energy\index{GRB!spectrum!break} $\epsilon_{\gamma b}$
of most GRBs 
observed by \BA\ detectors is in the range of
100~keV to 300~keV \cite{Brainerd99}. It may appear from
Eq.~(\ref{eq:E_gamma}) that the
clustering of break energies in this narrow energy range requires
fine tuning of fireball model parameters, which should naturally 
produce a much wider range of break energies. 
This is, however, not the case \cite{GSW01a}. Consider the dependence of 
$\epsilon_{\gamma b}$ on $\Gamma$. The strong $\Gamma$ dependence of
the pair production\index{GRB!fireball!pair production} optical depth, Eq.~(\ref{eq:taugg}), implies
that if the value of $\Gamma$ is smaller than the minimum value allowed 
by Eq.~(\ref{eq:Gmingg}), for which 
$\tau_{\gamma\gamma}(\epsilon_\gamma = 100{\rm MeV})\approx1$,
most of the high energy photons in the power-law distribution
produced by synchrotron emission\index{GRB!synchrotron!emission}, 
$dn_\gamma/d\epsilon_\gamma\propto\epsilon_\gamma^{-2}$, would be converted
to pairs. This would lead to high optical depth due to Thomson scattering\index{GRB!fireball!Thomson scattering} 
on $e^\pm$, and hence to strong suppression of the emitted flux
\cite{GSW01a}. For
fireball parameters such that 
$\tau_{\gamma\gamma}(\epsilon_\gamma = 100{\rm MeV})\approx1$, the break energy
implied by Eqs.~ (\ref{eq:E_gamma}) and  (\ref{eq:Gmingg}) is

\begin{equation}
\epsilon_{\gamma b}\approx1\xi_B^{1/2}\xi_e^{3/2}{L_{\gamma,52}^{1/6}
\over\Delta t_{-2}^{2/3}}~{\rm MeV}.
\label{eq:E_gamma-max}
\end{equation}

\noindent As explained in \S\ref{sec:eta}, shell Lorentz factors\index{GRB!fireball!Lorentz factor} cannot exceed
$\eta_*\simeq10^3$, for which break energies in the X-ray range,
$\epsilon_{\gamma b}\sim10$~keV, may be obtained. We note,
however, that the radiative flux would be strongly 
suppressed in this case too \cite{GSW01a}. 
If the typical $\Gamma$ of
radiation emitting shells is close to $\eta_*$, then the 
range of Lorentz factors of wind shells is narrow,
which implies that only a small fraction of wind kinetic
energy would be converted to internal energy which can be radiated from the
fireball.

Thus, the clustering of break energies at $\sim1$ MeV is naturally accounted 
for, provided that the variability time scale satisfies
$\Delta t\le10^{-2}$~s, which implies an upper limit on the source size,
since $\Delta t\ge r_0/c$ (see \cite{Levinson00,Ghisellini99} for
alternative explanations). We note, that 
a large fraction of bursts detected by \BA\ show variability
on the shortest resolved time scale, $\sim10$~ms \cite{Woods95}. 
In addition, a natural
consequence of the model is the existence of low luminosity
bursts with low, $1-10$ \keV, break energies \cite{GSW01a}. Such 
``X-ray bursts'' may have recently been identified \cite{FXTs}.

For internal collisions, the observed
$\gamma$-ray variability time, $\sim r_i/\Gamma^2 c\approx\Delta t$,
reflects the variability time of the underlying source, and the GRB
duration reflects the duration over which energy is emitted from the
source. Since the wind Lorentz factor is expected to fluctuate on
time scales ranging from the shortest variability time $r_0/c$ to the
wind duration $t_{\rm GRB}$, internal collisions will take place over a range
of radii, $r \sim \Gamma^2 r_0$ to $r \sim \Gamma^2 c t_{\rm GRB}$. 

\subsection{Afterglow Emission}
\label{sec:fireball-afterglow}

Let us consider the radiation emitted from the reverse shocks\index{GRB!shock!reverse} during the transition to self-similar\index{GRB!fireball!self-similar solution} expansion.
The characteristic electron
Lorentz factor (in the plasma rest frame)
is $\Gamma_m\simeq\xi_e(\Gamma/\Gamma_{\rm tr.})m_p/m_e$, where
the internal energy per proton in the shocked ejecta\index{GRB!ejecta!shocked}
is $\simeq (\Gamma/\Gamma_{\rm tr.})m_p c^2$. The energy density
$U$ is  $E \approx 4\pi r^2 c t \Gamma_{\rm tr.}^2 U$,
and the number of radiating electrons is $N_e\approx E/\Gamma m_p c^2$.
Using Eq.~(\ref{eq:Gamma_tr}) and $r = 4\Gamma_{\rm tr.}^2ct$, the 
characteristic (or peak) energy of
synchrotron photons\index{GRB!synchrotron!emission} (in the observer frame) is \cite{Draine00}

\begin{equation}
\epsilon_{\gamma m}
\approx\hbar\Gamma_{\rm tr.}\Gamma_m^2{eB\over m_e c} = 
2\xi_{e,-1}^2\xi_{B,-1}^{1/2}n_0^{1/2}
\Gamma_{2.5}^{2}~{\rm eV},
\label{eq:e_m}
\end{equation}

\noindent and the specific luminosity,
$L_\epsilon = dL/d\epsilon_\gamma$, at
$\epsilon_{\gamma m}$ is

\begin{eqnarray}
L_m&&\approx
(2\pi\hbar)^{-1}\Gamma_{\rm tr.}{e^3B\over m_e c^2}N_e\approx
10^{61}\xi_{B,-1}^{1/2}E_{53}^{5/4}t_1^{-3/4}
\Gamma_{2.5}^{-1}n_0^{1/4}{\rm\, s}^{-1},
\label{eq:L_m}
\end{eqnarray}

\noindent where $\xi_e = 0.1\xi_{e,-1}$, and $\xi_B = 0.1\xi_{B,-1}$.

Here too, we expect a power law energy distribution,
$dN_e/d\Gamma_e \propto \Gamma_e^{-p}$ for $\Gamma_e > \Gamma_m$, with $p\simeq2$.
Since the radiative cooling\index{GRB!synchrotron!cooling} time of electrons in the reverse shock\index{GRB!shock!reverse}
is long compared to the ejecta\index{GRB!ejecta} expansion time, 
the specific luminosity extends in this case to energy
$\epsilon_\gamma>\epsilon_{\gamma m}$ as
$L_\epsilon = L_m(\epsilon_\gamma/\epsilon_{\gamma m})^{-1/2}$,
up to photon energy
$\epsilon_{\gamma c}$. Here $\epsilon_{\gamma c}$ is
the characteristic synchrotron frequency of
electrons for which the synchrotron cooling time,
$6\pi m_e c/\sigma_T\gamma_e B^2$, is comparable to the ejecta (rest frame)
expansion time, $\sim \Gamma_{\rm tr.} t$. At energy
$\epsilon_\gamma > \epsilon_{\gamma c}$,

\begin{equation}
\epsilon_{\gamma c}\approx
0.1\xi_{B,-1}^{-3/2}n_0^{-1}E_{53}^{-1/2}t_1^{-1/2}
{\rm\, keV},
\label{eq:e_c}
\end{equation}

\noindent the spectrum steepens to $L_\epsilon\propto\epsilon_\gamma^{-1}$.

The shock driven into the \ISm\ continuously heats new gas, and
produces relativistic 
electrons that may produce the delayed afterglow\index{GRB!afterglow} radiation
observed on time scales $t \gg t_{\rm SS}$, typically of order
days to months. As the shock wave decelerates, the emission shifts to
lower frequency with time. Since afterglow emission on such long
time scale is extensively discussed in other chapters of this
volume, we do
not discuss in detail the theory of late-time afterglow emission.

\subsection{Open Questions: Magnetic Field and Electron Coupling}
\label{sec:openQs}

The emission of radiation in both the GRB and afterglow phases is
assumed to arise from synchrotron emission\index{GRB!synchrotron!emission} of shock\index{GRB!shock} accelerated electrons.
To match observations, the magnetic field\index{GRB!magnetic field} behind the shocks must be
close to equipartition\index{GRB!fireball!equipartition} and a significant fraction of
the internal shock energy must be carried by electrons, that is,
$\xi_B$ and $\xi_e$ should be close to unity, of order $10\%$. 
During the afterglow phase, shock compression of the existing \ISm\ field  
yields a field many orders of magnitude
smaller than needed. Thus, the magnetic
field\index{GRB!magnetic field} is most likely generated in, and by, the shock wave. A similar process
is likely necessary to generate the field required for synchrotron emission
during the GRB phase, \ie in the internal fireball shocks.
Although a magnetic field\index{GRB!magnetic field} 
close to equipartition at the base of the wind frozen into the fireball 
plasma may not be many orders of magnitude below equipartition\index{GRB!fireball!equipartition} during 
the internal shock phase, significant amplification is nevertheless required.
It is well known that near equipartition magnetic fields may be generated 
in collisionless shocks through the Weibel instability\index{Weibel instability} 
(see, \eg \cite{Kazimura98,Sagdeev66}). However, the field is generated on
microscopic, skin-depth, scale and is therefore expected to rapidly 
decay, unless its coherence length
grows to a macroscopic scale \cite{Gruzinov99,Gruzinov01}. The process
by which such scale increase is achieved is not understood, and probably
related to the process of particle acceleration \cite{Gruzinov01}.

In order to produce the observed spectrum during both afterglow and
GRB phases, electrons must be accelerated
in the collisionless shocks\index{GRB!shock!collisionless} to a power-law distribution,  
$dn_e/d\Gamma_e \propto \Gamma_e^{-p}$ with $p\simeq2$. As mentioned in 
\S\ref{sec:fireball-gammas}, such distribution is expected in
the internal shocks\index{GRB!shock!internal}, which are mildly
relativistic\index{GRB!shock!relativistic}. Recent numeric and analytic calculations of particle 
acceleration via the first order Fermi mechanism\index{Fermi acceleration} in relativistic shocks
show that similar spectral indices, $p\approx2.2$, are obtained
for highly-relativistic shocks as well \cite{BnO98,Kirk00}. The derivation
of electron spectral indices is based, in both the non-relativistic
and relativistic cases, on a phenomenological description of
electron scattering and, therefore, does not provide a complete basic principle
description of the process. In particular, these calculations do not
allow one to determine the fraction of energy $\xi_e$ carried by
electrons.

\section{The Fireball Model: Key Afterglow Implications}
\label{sec:fireball-AG}

The following 
point should be clarified in the context of afterglow observations -- the distribution of GRB durations is bimodal, with broad peaks at
$t_{\rm GRB} \sim 0.2$~s and $t_{\rm GRB} \sim 20$~s \cite{Fishman}. The
majority of bursts belong to the long duration\index{GRB!duration!long}, $t_{\rm GRB} \sim 20$~s,
class. The detection
of afterglow emission was made possible thanks to the accurate GRB
positions provided on hour time scale by the 
\B\ satellite \cite{Costa97-EW}. Since the detectors on board
this satellite trigger only on long bursts, afterglow observations are
not available for the smaller population of short\index{GRB!duration!short}, $t_{\rm GRB}\sim 0.2$~s,
bursts. Thus, while the discussion of the fireball model\index{GRB!fireball} presented in
\S\S\ref{sec:fireball-hyd} and \ref{sec:fireball-rad}, 
based on $\gamma$-ray observations and on simple
phenomenological arguments, applies to both long and short 
duration bursts, 
the discussion below of afterglow\index{GRB!afterglow} observations applies to long duration
bursts only. It should, therefore, be kept in mind that short duration bursts
may constitute a different class of GRBs which, for example,
may be produced by a 
different class of progenitors and may have a different redshift distribution
than the long duration bursts.

Afterglow observations led to the confirmation, 
as mentioned in \S1, of 
the cosmological origin of GRBs \cite{AG_ex_review},
and supported \cite{Waxman97a,Wijers97-EW} standard model
predictions \cite{Katz94,MnR97,Rhoads93,Vietri97a}
of afterglows that result from 
synchrotron emission\index{GRB!synchrotron!emission} by electrons accelerated to
high energy in the highly relativistic shock driven by the fireball\index{GRB!fireball} into
its surrounding gas\index{GRB!circumburst medium}. As discussed in separate chapters of this volume,
both the spectral and temporal behavior of afterglow emission are in
general agreement with model predictions.

Since afterglow emission results from the interaction of the fireball
with ambient medium, it does not provide direct information on the 
evolution of the fireball at the earlier stage during which the
GRB is produced. Nevertheless, afterglow observations may be used to
indirectly test underlying model assumptions and constrain model parameters
relevant for this earlier stage. We describe below several
afterglow observations which have important implications for the GRB
phase of the model.

\subsection{Fireball Size and Relativistic Expansion}
\label{sec:scintillation}

Radio\index{GRB!radio!emission} observations of \GRB{970508} allowed a direct determination of
the fireball size and a direct confirmation of its relativistic expansion.
As explained in \S\ref{sec:Rel-expansion}, radiation seen by a distant observer
originates from a conical section of the fireball around the source-observer 
line of sight, with opening angle $\sim1/\Gamma$, and photons which are emitted
from the edge of the cone are delayed compared to those emitted along the
line of sight, by $r/2\Gamma^2c$. Thus, the 
apparent radius of the emitting cone is $R = r/\Gamma(r) = 2\Gamma(r) ct$
where $r$ and $t$ are related by $t = r/2\Gamma(r)^2c$.
(A detailed calculation of fireball emission
introduces only a small correction, $R = 1.9\Gamma(r) ct$ \cite{WAG-ring}.
Using Eq.~\ref{eq:Gamma_BM}, we find that the apparent size of the fireball
during its self-similar expansion\index{GRB!fireball!self-similar solution} into the surrounding medium is 
given by 

\begin{equation}
R = 0.8\times10^{17}\left({1+z\over2}\right)^{-5/8}\left({E_{52}\over n_0}
\right)^{1/8}\left({t\over1\rm week}\right)^{5/8}{\rm cm}.
\label{eq:R}
\end{equation}

\noindent We have chosen here the normalization $E = 10^{52}E_{52}$ erg since this is the
energy inferred for \GRB{970508} \cite{Waxman97b,WKF98} (note, however,
the very weak dependence on $E$ and $n$). The factor 
$1+z$ appears due to the redshift between source and observer time intervals.

\begin{figure}[t]
\begin{center}
\includegraphics[angle = 270,width = 4.5in]{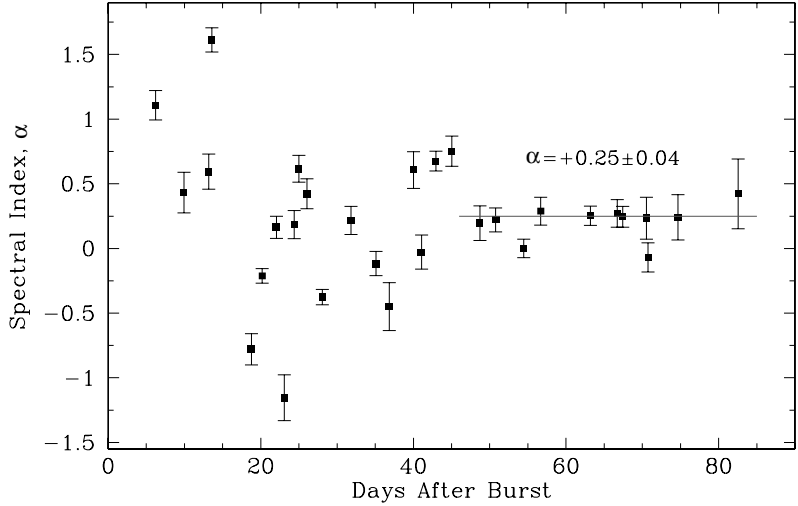}
\end{center}
\caption{The ratio between the \GRB{970508} afterglow
radio fluxes at 4.86 GHz and 8.46 GHz,
$\alpha \equiv \log[f_\nu(4.86{\rm GHz})/f_\nu(8.46{\rm GHz})]/\log(4.86/8.46)$,
is shown as a function of time following the burst (modified from
\cite{FWK00}).
The rapid variations at early times are due to narrow band diffractive 
scintillation, and their quenching at late times is due to the expansion 
of the source beyond a critical size given by Eq.~(\ref{eq:Rdiff}). 
This is a direct confirmation of model
predictions, according to which highly relativistic plasma ejection
is responsible for the observed radiation.}
\label{fig:scintillation}
\end{figure}

Scattering by irregularities in the local interstellar medium \ISm\ may 
modulate the observed fireball radio flux\index{GRB!radio!scintillation} \cite{Goodman97-EW}. If scattering
produces multiple images of the source, interference between the multiple
images may produce a diffraction pattern, leading to strong variations of 
the flux as the observer moves through the pattern. In order for the 
diffraction patterns produced by different points on the 
source to be similar, so that the pattern is not smoothed out due
to large source size, the apparent size of a source a redshift $z = 1$
must satisfy \cite{WKF98}

\begin{equation}
R<R_{\rm sc.}\approx10^{17}{\nu_{10}^{6/5}\over h_{75}}
\left({SM\over10^{-3.5}{\rm m}^{-20/3}{\rm kpc}}\right)^{-3/5}\,{\rm cm},
\label{eq:Rdiff}
\end{equation}

\noindent where $\nu = 10\nu_{10}$ GHz, $h_{75}$ is the Hubble Constant in units of
75 \kmsMpc, and the scattering measure, $SM$, a measure of the
strength of the electron density fluctuations, is normalized to 
its characteristic Galactic value.

Comparing Eqs.~ (\ref{eq:R}) and (\ref{eq:Rdiff}) we find that, on time scale 
of weeks, the apparent fireball size\index{GRB!fireball!size} is comparable to the maximum size for which
diffractive scintillation\index{GRB!radio!scintillation} is possible. On shorter time scales, therefore,
strong modulation of the radio flux is expected, while on longer time scales
we expect diffractive scintillation to be quenched. This is exactly
what had been observed for \GRB{970508}, as demonstrated in 
Fig.~\ref{fig:scintillation}. Observations are therefore in agreement with 
fireball model predictions: They imply a source size consistent with
model prediction, Eq.~(\ref{eq:R}) which, in particular, imply expansion 
at a speed comparable to that of light. 

\subsection{The Nature of the Fireball Plasma}
\label{sec:UV-flash}

Due to present technical limitations of the experiments, 
afterglow\index{GRB!afterglow} radiation is observed in most cases only on time scale $\gg 10$~s.
In one case, however, 
that of \GRB{990123}, optical emission\index{GRB!optical!emission} was detected on $\sim10$~s
time scale \cite{Akerlof99}. The most natural explanation of the 
observed optical radiation is synchrotron emission\index{GRB!synchrotron!emission} from electrons accelerated
to high energy in the reverse shocks\index{GRB!shock!reverse} driven into fireball ejecta\index{GRB!ejecta} at the
onset of interaction with the surrounding medium\index{GRB!circumburst medium} \cite{MR_0123,SP_0123},
as explained in \S\ref{sec:fireball-afterglow}.
This observation provides, therefore, direct constraints on the fireball
ejecta plasma. First, it provides
strong support for one of the underlying assumptions of the 
dissipative fireball scenario described in 
\S\ref{sec:fireball-evolution}, that the energy is carried from the 
underlying source in the form of proton kinetic energy. This is due to the
fact that the observed radiation is well accounted for in a model
where a shock\index{GRB!shock} propagates
into fireball plasma composed of protons\index{GRB!fireball!protons} and electrons\index{GRB!fireball!electrons} (rather than,
\eg a pair plasma). Second, 
comparison of the observed flux with model predictions from  Eqs.~ 
(\ref{eq:e_m}) and (\ref{eq:L_m}) implies
$\xi_e\sim\xi_B\sim10^{-1}$. 

\subsection{Gamma-Ray Energy and GRB Rate}
\label{sec:z-distribution}

Following the determination of GRB redshifts, it is now clear that most
GRB sources lie within the redshift range $z \sim 0.5$ to $z \sim 2$, with 
some bursts observed at $z>3$. For the average GRB $\gamma$-ray fluence,
$1.2 \times 10^{-5}$ \ergcm\ in the 20~keV to 2~MeV band, 
this implies characteristic isotropic $\gamma$-ray energy and luminosity\index{GRB!explosion energy}
$E_\gamma\sim10^{53}$ erg and $L_\gamma \sim 10^{52}$ \ergsec\
(Here, we assume a flat universe
with $\Omega = 0.3$, $\Lambda = 0.7$, and $H_0 = 65$ \kmsMpc. 
These estimates are consistent with more detailed analyses of the 
GRB luminosity function and redshift distribution. Mao and Mo \cite{MnM98}, \eg find, for the cosmological parameters we use, 
a median GRB energy of $\approx0.6 \times 10^{53}$ erg in the $50-300$ \keV\ band, corresponding to a median GRB energy 
of $\approx2 \times 10^{53}$ erg in the 20 \keV\ to 2~MeV band.

Since most observed GRB sources lie within the redshift range $z \sim 0.5-2$, observations
essentially determine the GRB rate\index{GRB!rate} per unit volume
at $z\sim1$. The observed rate of $10^3$ yr$^{-1}$ implies 
$R_{\rm GRB}(z = 1) \approx 3~{\rm Gpc}^{-3}~{\rm yr}^{-1}$
(for $\Omega = 0.3$, $\Lambda = 0.7$, and $H_0 = 65$ \kmsMpc. 
The present, $z = 0$, rate is less well constrained since available data
are consistent with both no evolution of GRB rate\index{GRB!rate!evolution} with redshift and 
with strong evolution (following, \eg the star formation rate), in which $R_{\rm GRB}(z = 1)/R_{\rm GRB}(z = 0) \sim10$
\cite{GRB_z2,GRB_z1}.
A detailed analysis by, \eg Schmidt \cite{Schmidt01} leads, assuming $R_{\rm GRB}$ is 
proportional to the star formation rate, to 
$R_{\rm GRB}(z = 0) \sim0.5~{\rm Gpc}^{-3}~{\rm yr}^{-1}$.

If fireballs are conical jets of solid angle 
$\Delta\Omega$ then, clearly, the
total $\gamma$-ray energy is smaller by a factor
$\Delta\Omega/4\pi$ than the isotropic energy, and the GRB rate is larger
by the same factor.

\subsection{Fireball Geometry}
\label{sec:jets}

Afterglow observations suggest that at least some GRBs are conical jets,
of opening angle $\theta \sim 10^{-1}$ corresponding to a solid angle 
$\Delta\Omega\sim10^{-2}$ \cite{Tirado-jet,Kulkarni-jet,SPH-jet}.
As explained in \S\ref{sec:fireball-geometry}, the discussion in 
\S\ref{sec:fireball-hyd} and \S\ref{sec:fireball-rad} is limited to the
stage where the wind is ultra-relativistic, 
$\Gamma\sim300$, prior to significant fireball deceleration\index{GRB!fireball!deceleration} by the 
surrounding medium\index{GRB!circumburst medium}, and is hence equally valid for 
both a spherical and a jet-like fireball. A jet\index{GRB!jet} like geometry has, of course,
profound implications for the underlying progenitor 
(see \S\ref{sec:progenitors}): it implies that
the underlying source must produce a 
collimated outflow and, if $\theta\sim10^{-1}$ is indeed typical, it implies
that the characteristic $\gamma$-ray energy emitted by the source is
$\approx10^{51}$ erg rather than $\approx10^{53}$ erg implied 
by the assumption of isotropy.

\subsection{Fireball $\gamma$-Ray Efficiency}
\label{sec:efficiency}

Afterglow observations imply that a significant fraction
of the energy initially carried by the fireball\index{GRB!fireball!efficiency} is converted into 
$\gamma$-rays, \ie that the observed $\gamma$-ray energy provides a rough
estimate of the total fireball energy. This has been demonstrated for
one case, that of \GRB{970508}, by a comparison of the total fireball energy
derived from long term radio\index{GRB!radio!emission} observations with the energy emitted in 
$\gamma$-rays \cite{FWK00,WKF98}, and for a large number of bursts
by a comparison of observed $\gamma$-ray energy with 
the total fireball energy estimate based on X-ray afterglow data 
\cite{Freedman}. Freedman and Waxman \cite{Freedman} demonstrated that  
a single measurement of the
X-ray afterglow\index{GRB!X-ray!emission} flux on the time scale of a day
provides a robust estimate of the
fireball energy per unit solid angle, $\varepsilon$, averaged
over a conical section of the fireball of opening angle $\theta\sim0.1$.
Applying their analysis to \B\ afterglow data they demonstrated that
the ratio of observed $\gamma$-ray to total fireball energy
per unit solid angle, $\varepsilon_\gamma/\varepsilon$,
is of order unity.

The inferred high radiative efficiency implies that 
a significant fraction of the wind kinetic energy must be converted to 
internal energy in internal shocks, and that electrons must carry a 
significant fraction of the internal energy, \ie that $\xi_e$ should
be close to unity. We have already shown in \S\ref{sec:fireball-gammas}
and \S\ref{sec:fireball-break}, following 
\cite{GSW01a}, that efficient conversion of kinetic
to internal energy and $\xi_e$ values close to unity may naturally lead
to $\sim1$~MeV spectral break\index{GRB!spectrum!break} energies, in accordance with observations.

\section{Progenitors Clues}
\label{sec:progenitors}

The observational constraints that provide the main hints regarding the
nature of the underlying GRB progenitors\index{GRB!progenitor} are the rapid, $\sim1$~ms,
$\gamma$-ray signal variability, and the total
$\gamma$-ray energy emitted by the source, 
$\approx10^{53}(\Delta\Omega/4\pi)$ erg. The rapid variability implies a 
compact object, of size smaller than a light millisecond, $\sim10^7$~cm,
and mass smaller than $\sim30$ \Msun\ (the mass of a black hole\index{black hole} with
$\sim10^7$~cm Schwarzschild radius\index{Schwarzschild radius}). The energy released in 
$\gamma$-rays corresponds to a $0.1(\Delta\Omega/4\pi)$ \Msun\ rest mass
energy. The most natural way for triggering the GRB event is therefore
the accretion of a fraction of a solar mass onto a (several) solar
mass black hole (or possibly a neutron\index{star!neutron} star). 
The dynamical time of such a source, comparable to
its light crossing time, is sufficiently short to account
for the observed rapid variability and, if significant fraction of the
gravitational binding energy release is converted to $\gamma$-rays,
the source will meet the observed energy requirements. 

Most cosmological
GRB models therefore have at their basis gravitational collapse of a
several solar-mass progenitor to a black hole\index{black hole}. Within
the context of the fireball\index{GRB!fireball} model, the observed variability is determined
by the dynamical time of the source, which determines the variability
in the ejected wind properties, while the GRB duration, $\ge10$~s
for long bursts, reflects the
wind duration, \ie the duration over which energy is extracted from
the source. The characteristic time for the gravitational collapse is of the
order of the dynamical time, \ie much shorter than the wind duration.  
Most models therefore assume that, following collapse and black hole
formation, some fraction of the progenitor mass forms an accretion disk
which powers the wind through gradual mass accretion. The 
characteristic time scale for accretion is set by the disk viscosity, which
is uncertain and assumed to correspond to the observed GRB duration.

Progenitor models differ in the scenario for black hole\index{black hole} formation, and
in the process assumed to convert disk energy into relativistic outflow.
The two leading progenitor scenarios are, at present, collapses of
massive stars\index{GRB!progenitor!collapsar} \cite{BohdanHN,Woosley93-EW}, and mergers of compact objects\index{GRB!progenitor!binary merger}
\cite{Goodman86-EW,Bohdan86}. In the former case, the progenitor is a massive
rotating star, \eg a $\sim15$ \Msun\ helium star evolved (by mass-loss) from
a $\sim30$ \Msun\ main sequence star. The collapse of the progenitor's
$\sim2$ \Msun\ iron core leads to the formation of a black hole surrounded by an
accretion disk composed of mantle plasma \cite{Woosley93-EW}. 
In the latter case, the merger of
two neutron stars leads to the formation of a black hole surrounded by 
a disk produced by neutron star disruption during the merger process 
(see, \eg \cite{Eichler92}). A similar scenario involves neutron star
disruption during a neutron star--black hole merger (see, \eg \cite{Lattimer74}). 

Two types of processes are widely considered for the extraction of 
disk energy: neutrino emission and magneto-hydrodynamic (MHD) processes. 
The viscous
dissipation of energy, driving mass accretion, heats the dense disk plasma
leading to the emission of thermal neutrinos\index{GRB!neutrinos} of all flavors. Neutrino 
annihilation along the rotation axis in the vicinity of the black hole
may then produce an electron-positron pair plasma fireball\index{GRB!fireball!pair production}. The fraction
of rest mass which is dissipated during accretion is typically of order
10\% (the specific energy at the last stable orbit 
of a non-rotating black-hole, at three Schwarzschild radii\index{Schwarzschild radius}, is $c^2/6$). Accretion
of 0.1 \Msun\ over a second may therefore lead to a neutrino luminosity
of $\approx10^{52.5}$ \ergsec, of which $\approx1\%$ would be
deposited by annihilation to drive a fireball. The resulting wind luminosity,
$\sim10^{50}$ \ergsec, may be too low to drive a spherical fireball,
but may be sufficient if the fireball is collimated into 
$\Delta\Omega/4\pi\sim10^{-2}$.  

If equipartition\index{GRB!fireball!equipartition} magnetic fields\index{GRB!magnetic field}, $\sim10^{15}$~G, are built in  
the disk (\eg by convective motion) the dissipated energy may be extracted
electromagnetically from the disk. Although the process by which such a strong
field is generated, as well as the details of the energy extraction mechanism
in the presence of such field are not understood, there is evidence from
various astrophysical systems (active galactic nuclei\index{galaxy!AGN} and  micro-quasars\index{galaxy!quasar!micro}, see, \eg \cite{Livio-jets99}), for the formation of MHD driven jets\index{GRB!jet} which carry
$\sim10\%$ of the disk binding energy. The presence of equipartition fields
may also allow the extraction of  energy directly, \eg via the Blandford-Znajek
mechanism\index{Blandford-Znajek
mechanism} \cite{Znajek77}, from the rotating black hole \cite{Levinson93}. 
For rapid rotation, the available energy in this case is comparable to the 
collapsed rest mass. Thus, MHD processes are often invoked to drive a
relativistic wind with efficiency much higher than that estimated for 
a neutrino driven wind. 

We note in this context that 
a possible alternative to the above models may be the formation 
from stellar collapse of a fast rotating neutron\index{star!neutron} star with ultra-high 
magnetic field\index{GRB!magnetic field} \cite{Thompson94,Usov94}.
If a fast rotating, millisecond period, neutron star is produced by the
collapse with $\sim10^{15}$~G field, then the resulting
electromagnetic energy luminosity is sufficient to drive a GRB wind.

One problem which all models are facing is the baryon loading\index{GRB!fireball!baryons}. In order to
allow acceleration to the high Lorentz factors implied by observations, 
$\Gamma \sim 10^3$, the total mass entrained within the expanding plasma
must be smaller than $\sim10^{-4}$ \Msun, \ie the mass-loss rate\index{GRB!progenitor!mass-loss} should
be smaller than $\sim10^{-6}$ \Msunsec. The neutrino luminosity
is expected to drive mass-loss at a much higher rate. It is generally assumed
that mass flow towards the rotation axis is inhibited (\eg by high pressure 
of fireball plasma along the rotation axis), thus allowing the formation
of a sufficiently baryon free fireball collimated along the rotation axis.
In the case of a massive star progenitor\index{GRB!progenitor!collapsar}, the fireball jet\index{GRB!jet} is assumed to form
along the rotation axis of the star, where rapid rotation leads to lower
mantle and envelope density. The collimation of the fireball\index{GRB!fireball} may be due
in this case to the presence of a low density funnel along the rotation
axis, and the resulting jet must penetrate through the stellar mantle
and envelope in order to allow the production of an observable GRB.
Recent numerical and analytical calculations of the propagation
of high entropy jets through stellar progenitors indicate that such
a scenario may be viable \cite{Aloy00,Fryer98,MnR-funnel01}. 

Afterglow\index{GRB!afterglow} observations provide several hints which indicate that
long duration GRBs\index{GRB!duration!long}, at least those for which afterglows have been detected, are associated
with massive star progenitors.  The location of GRBs within host galaxies,
the presences of iron lines, and the evidence for a supernova
association\index{GRB!supernova connection} all imply massive star progenitors. Since this issue is discussed in detail in
other chapters in this volume, we address it only briefly here. 

Most GRB afterglows
are localized within the optical image of a host galaxy 
\cite{Bloom-location01}.
This is in disagreement with simple analyses of the neutron star merger\index{GRB!progenitor!n*-n* merger} scenario which predict that the high velocity of such binaries should carry many of them
outside of the host prior to merger. This result is, however, uncertain since
it depends on model parameters which are only poorly constrained, such as the distribution of initial binary separations. Evidence for the presence
of iron lines has been found in X-ray data for two bursts
\cite{Amati-Fe00,Piro-Fe00}. While 
the presence of iron lines strongly suggests a massive stellar progenitor\index{GRB!progenitor!collapsar}, 
as it indicates the presence of iron enriched
environment, the confidence level of their detection is moderate.
There is evidence in three cases that a supernova may be 
associated with the
GRB \cite{Galama-SN98,Paradijs-SN99}. The evidence is, however, not yet
conclusive (see, \eg \cite{Draine00}). 
Finally, the synchrotron emission\index{GRB!synchrotron!emission} produced by a shock driven by the
fireball\index{GRB!fireball} into its surrounding medium\index{GRB!circumburst medium} depends on the density of the 
ambient medium. Thus, the temporal and spectral dependence of this
afterglow emission may distinguish between the high density environment 
characteristic of a massive stellar wind, expected to exist in the
case of a massive stellar progenitor, and the low density \ISm\ expected, \eg in merger scenarios. Present observations
are not yet conclusive, since data on time scales much shorter than one
day are required to distinguish between the two cases \cite{Livio00-EW}.

It is clear from the above discussion, that future afterglow\index{GRB!afterglow}
observations providing more detailed information on the burst
environment\index{GRB!circumburst medium} and location will play a crucial role in placing 
stringent constraints on progenitor models. In addition to the
points discussed in the previous paragraph, the destruction of
dust \cite{Draine02,FKR01,Reichart01,Draine00} and time dependence
of atomic, ionic \cite{Boettcher99,Perna98}
and molecular H$_2$ \cite{Draine02,Draine00} lines due to photoionization
may be detectable for a burst in a molecular cloud environment
characteristic of star forming regions.

\section{High Energy Protons and Neutrinos from GRB Fireballs}
\label{sec:UHECR}

\subsection{Fermi Acceleration in GRBs}
\label{sec:fermi}

In the fireball model, the observed radiation is produced, both during
the GRB and the afterglow, by synchrotron emission from shock accelerated
electrons. In the region where electrons\index{GRB!fireball!electrons} are accelerated, 
protons\index{GRB!fireball!protons} are also expected to be
shock\index{GRB!shock} accelerated. This is similar to what is thought to occur in supernovae 
remnant shocks, where synchrotron radiation of accelerated electrons is the
likely source of non-thermal X-rays (recent \ASCA\ observations give evidence
for acceleration of electrons in the remnant of SN1006\index{supernova remnant!individual!SN1006} to $10^{14}$ eV \cite{SN1006}), and where shock acceleration of protons is believed to
produce cosmic rays\index{cosmic-rays} with energy extending to $\sim10^{15}{\rm eV}$ (see, \eg \cite{Blandford87} for a review). Thus, it is likely that protons, as well
as electrons, are accelerated to high energy within GRB fireballs.

We consider proton Fermi acceleration\index{Fermi acceleration} in fireball internal\index{GRB!shock!internal} and
reverse\index{GRB!shock!reverse} shocks (see \S\ref{sec:fireball-evolution} and 
\S\ref{sec:fireball-interaction} respectively). 
Since these shocks are mildly relativistic, 
with Lorentz factors\index{GRB!shock!Lorentz factor} $\Gamma_i-1\sim1$ in the wind frame 
(see \S\S\ref{sec:fireball-interaction} and \ref{sec:fireball-gammas}),
the predicted energy distribution of accelerated
protons is \cite{Bell78,Blandford87} $dn_p/d\epsilon_p\propto \epsilon_p^{-2}$,
similar to the
electron energy spectrum inferred from the observed photon spectrum.

Two constraints must be satisfied by
fireball wind parameters in order to allow proton acceleration to
$\epsilon_p>10^{20}$~eV in internal and reverse shocks.
First, the proton acceleration time, 
$t_a\sim R_L/c$ where $R_L$ is the proton Larmor radius\index{Larmor radius},
must be smaller than the wind expansion time \cite{MnU95,Vietri95,W95a}, 
$t_d\sim r_i/\Gamma c$ (in the 
wind frame). This constraint sets a lower limit to the 
magnetic field\index{GRB!magnetic field} carried by the wind, which may be expressed as \cite{W95a}:

\begin{equation}
\xi_B/\xi_e>0.02\Gamma_{2.5}^2 \epsilon_{p,20}^2L_{\gamma,52}^{-1}.
\label{eq:xi_B}
\end{equation}

\noindent Here, $\epsilon_p = 10^{20}\epsilon_{p,20}$~eV. Recall that
$\xi_B$ is the 
fraction of the wind energy density which is carried by magnetic field\index{GRB!magnetic field},
$4\pi r^2 c\Gamma^2 (B^2/8\pi) = \xi_B L$, 
and $\xi_e$ is the fraction of wind energy carried by shock
accelerated electrons. Since the electron energy is lost radiatively,
$L_\gamma\approx\xi_e L$. 

The second constraint that should be satisfied is that the 
proton synchrotron loss time\index{GRB!synchrotron!losses} must exceed $t_a$, setting an upper 
limit to the magnetic field\index{GRB!magnetic field}. The latter constraint may be satisfied 
simultaneously with the lower limit to the magnetic field\index{GRB!magnetic field},
Eq.~(\ref{eq:xi_B}), provided \cite{W95a}

\begin{equation}
\Gamma>130 \epsilon_{p,20}^{3/4}\Delta t^{-1/4}_{-2}.
\label{eq:G_min}
\end{equation}

The constraints that must be satisfied to allow acceleration of protons 
to energy $>10^{20}$~eV are remarkably similar to those inferred from
$\gamma$-ray observations: $\Gamma>100$ is implied by observed
$\gamma$-ray spectra (see \S\S\ref{sec:Rel-expansion} and \ref{sec:eta}), and 
magnetic field\index{GRB!magnetic field} close to equipartition\index{GRB!fireball!equipartition}, $\xi_B\sim1$, is required
in order for electron synchrotron emission\index{GRB!synchrotron!emission} to account for the observed
radiation (see \S\ref{sec:fireball-gammas}). 

It has recently been claimed \cite{Gallant98} that the
conditions at the external shock\index{GRB!shock!external} driven by the fireball into the ambient
gas\index{GRB!circumburst medium} are not likely to allow proton acceleration to ultra-high energy. 
Regardless of the validity of this claim, it
is irrelevant for the acceleration in internal shocks,
the scenario considered for \UHE\ production in GRBs \cite{Vietri95,W95a}. Moreover, it is not at all clear that \UHE s cannot
be produced at the external shock, since the magnetic field\index{GRB!magnetic field} may be 
amplified ahead
of the shock by the streaming of high energy particles.
For discussion of high energy proton production in the external shock and
its possible implications see Dermer \cite{Dermer00}.

\subsection{\UHE\ Flux and Spectrum}
\label{sec:CRflux}

The local ($z = 0$) energy production rate in $\gamma$-rays by GRBs
is roughly given by the product of the characteristic GRB 
$\gamma$-ray energy, $E\approx2\times10^{53}$ erg, and the 
local GRB rate\index{GRB!rate}. Under the assumption that the GRB rate evolution\index{GRB!rate!evolution} 
is similar to the star-formation rate\index{star!formation rate} evolution, 
the local GRB rate is $\sim0.5~{\rm Gpc}^{-3}~{\rm yr}^{-1}$ 
\cite{Schmidt01} (see \S\ref{sec:z-distribution}), implying
a local $\gamma$-ray energy generation rate of 
$\approx10^{44}~{\rm erg}~{\rm Mpc}^{-3}~{\rm yr}^{-1}$.
The energy observed in $\gamma$-rays reflects the fireball\index{GRB!fireball}
energy in accelerated electrons\index{GRB!fireball!electrons}. Thus, if 
accelerated electrons and protons\index{GRB!fireball!protons} carry similar energy 
(as indicated by afterglow observations \cite{Freedman} (see, however, \cite {Totani00})
then the GRB production rate of high energy protons is

\begin{equation}
\epsilon_p^2 (d\dot n_p/d\epsilon_p)_{z = 0}\approx 
10^{44}~{\rm erg}~{\rm Mpc}^{-3}~{\rm yr}^{-1}.
\label{eq:cr_rate}
\end{equation}

\begin{figure}[t]
\begin{center}
\includegraphics[height = 3.5in]{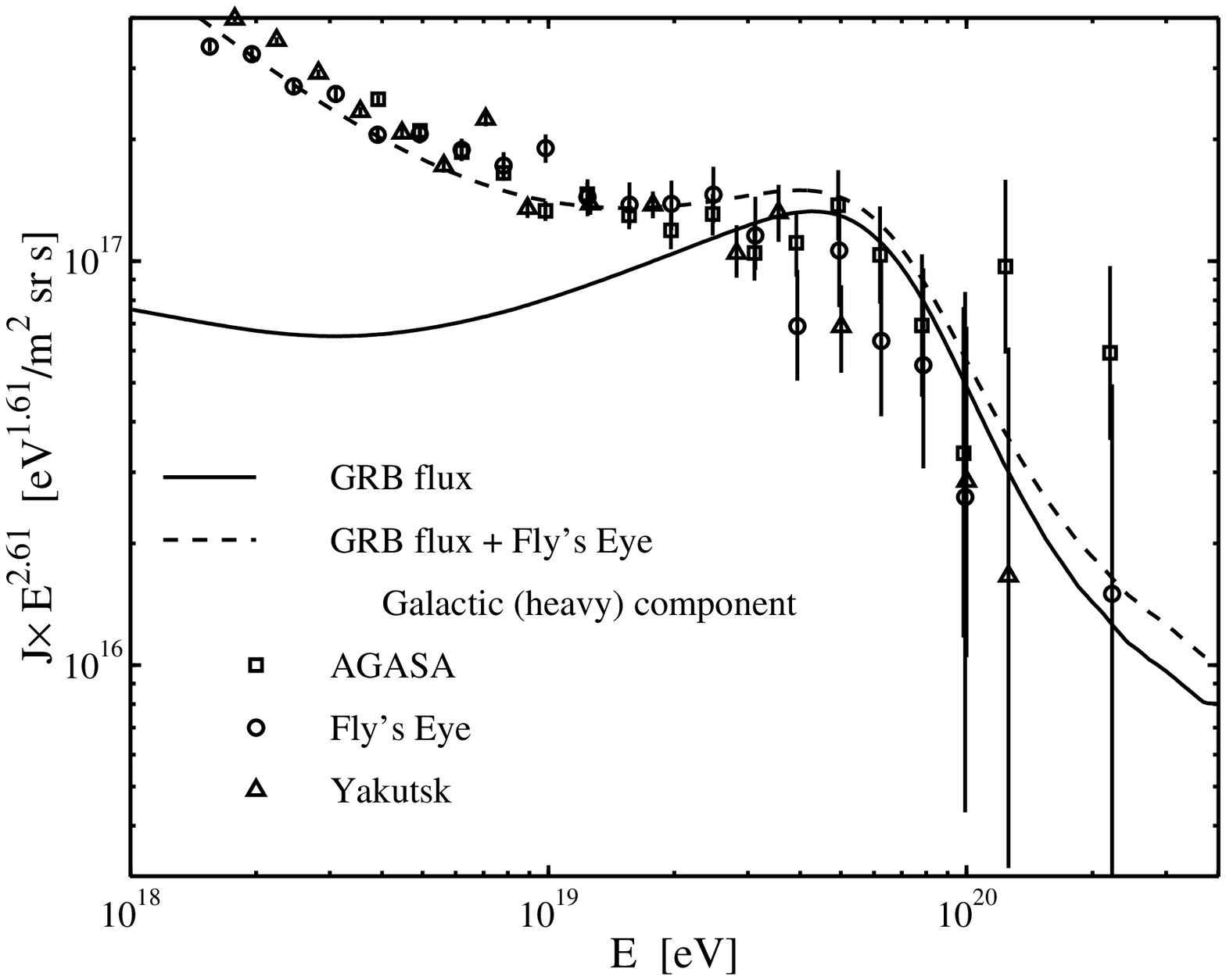}
\end{center}
\caption{The \UHE\ flux expected in a cosmological model 
where high-energy protons 
are produced at a rate $(\epsilon_p^2 d\dot n_p/d\epsilon_p)_{z = 0} = 
0.8\times10^{44}$ erg Mpc$^{-3}$ yr$^{-1}$ as predicted in the GRB model 
(Eq.~\ref{eq:cr_rate}, solid line), compared to the \HIRES\ \cite{Bird94}, \YAK\ \cite{Yakutsk} and \AGASA\ \cite{Takeda98} data. 
$1\sigma$ flux error bars are shown. 
The dashed line is the sum of the GRB model flux and the \HiRes\ fit 
(with normalization increased by 25\%) to
the Galactic heavy nuclei component \cite{Bird94}, $J_G \propto E^{-3.5}$,
which dominates below $\sim10^{19}$~eV. }
\label{fig:cr-flux}
\end{figure}

In Fig.~\ref{fig:cr-flux} we compare the observed \UHE\ spectrum
with that predicted by the GRB model. 
The generation rate (Eq.~\ref{eq:cr_rate}) of high energy protons
is remarkably similar to that 
required to account for the flux of $>10^{19}$~eV cosmic-rays\index{cosmic-rays}.
The flux at lower energies is most likely dominated by heavy nuclei
of Galactic origin \cite{Bird94}, as indicated by 
the flattening of the spectrum at $\approx10^{19}$~eV and by the evidence
for a change in composition at this energy 
\cite{Bird93,Bird94,composition2,composition1,Watson91}.

The suppression of model flux above $10^{19.7}$~eV is  
due to energy loss of high energy protons
in interaction with the microwave background, \ie to the ``GZK cutoff\index{GZK cutoff}'' \cite{GZK1,GZK2}. The available data do not allow determination of
the existence (or absence) of the ``cutoff''
with high confidence. The \AGASA\ results show an excess ($\sim2.5\sigma$ confidence level) of events compared to model
predictions above $10^{20}$ eV.  This excess is not confirmed,
however, by other experiments. Moreover, since the $10^{20}$ eV flux is dominated by sources at distances $<100\ {\rm Mpc}$, over
which the distribution of known astrophysical systems
(\eg galaxies and clusters of galaxies) is inhomogeneous,
 significant deviations from model predictions presented
in Fig.~\ref{fig:scintillation} for a uniform source distribution are expected \cite{W95b}. Clustering of cosmic-ray sources leads
to a standard deviation, $\sigma$, in the expected number, $N$, of 
events above $10^{20}$ eV given by \cite{CR_clustering}

\begin{equation} 
\sigma /N = 0.9(d_0/10 {\rm Mpc})^{0.9},
\end{equation}

\noindent where $d_0$ is the unknown scale
length of the source correlation function and $d_0\sim10$ Mpc 
for field galaxies.

Although the rate of GRBs\index{GRB!rate} out to a distance of 100~Mpc from
Earth, the maximum distance traveled by $>10^{20}$~eV protons, is
in the range of $10^{-2}$ to $10^{-3}$ \yr, the number of 
different GRBs contributing to the flux of $>10^{20}$~eV protons at any given
time may be large. This is due to the dispersion, $\Delta t$, in proton
arrival time, which is expected due to deflection by
intergalactic magnetic fields\index{GRB!magnetic field} and may 
be as large as $10^{7}$ \yr.  This implies that 
the number of sources contributing to the 
flux at any given time may be as large as \cite{W95a} 
$\sim\Delta t \times 10^{-3}$ \yr~$\sim 10^4$.

\subsection{Neutrino Production}
\label{sec:neutirnos}

A burst of $\sim10^{14}$ eV
neutrinos\index{GRB!neutrinos} accompanying observed $\gamma$-rays  
is a natural consequence of the conventional fireball\index{GRB!fireball} scenario \cite{WnB97}.
The neutrinos are produced by $\pi^+$ created in interactions
between fireball $\gamma$-rays\index{GRB!gamma-ray!emission} and
accelerated protons\index{GRB!fireball!protons}.
The key relation
is between the observed photon energy, $\epsilon_\gamma$,
and the accelerated proton's energy, $\epsilon_p$,
at the photo-meson threshold  of the $\Delta$-resonance.
In the observer frame,

\begin{equation}
\epsilon_\gamma ~ \epsilon_{p} = 0.2~\Gamma^2~{\rm GeV^2}.
\label{eq:keyrelation}
\end{equation}

\noindent For $\Gamma \approx 300$ and $\epsilon_\gamma = 1$~MeV,
we see that characteristic proton energies
$\sim 10^{16}$~eV are required to produce pions\index{GRB!fireball!pions}.
The pion typically carries $\approx20\%$ of the interacting proton energy,
and this energy is roughly equally distributed between the leptons in the
decay $\pi^+\rightarrow\mu^++\nu_\mu
\rightarrow e^++\nu_e+\overline\nu_\mu+\nu_\mu$. Thus, proton interaction with
fireball $\gamma$-rays is expected to 
produce $\sim 10^{14}$~eV neutrinos.

The fraction $f_\pi(\epsilon_p)$
of proton energy lost to pion production is determined by the
number density of photons in the dissipation region and is 
$\approx20\%$ at high proton energy for fireball wind
parameters implied by $\gamma$-ray observations \cite{GSW01b,WnB97}.
Assuming that GRBs produce high energy protons
at a rate given by Eq.~(\ref{eq:cr_rate}), the intensity of high energy
neutrinos is \cite{WnB97}

\begin{equation}
\epsilon_\nu^2\Phi_{\nu_x}\approx
10^{-9}\left({f_\pi\over0.2}\right)\min\left(1,{\epsilon_\nu\over10^{14}
{\rm eV}}\right)~{\rm GeV}~{\rm cm}^{-2}~{\rm sr}^{-1}~{\rm s}^{-1}.
\label{eq:JGRB}
\end{equation}

\noindent Here, $\nu_x$ stands for $\nu_\mu$, $\bar\nu_\mu$ or $\nu_e$.
The neutrino\index{GRB!neutrinos} flux of Eq.~(\ref{eq:JGRB}) is suppressed 
at high energy, $>10^{16}$~eV, due to synchrotron energy loss of pions and 
muons\index{GRB!fireball!muons} \cite{RnM98,WnB97}.

During the transition to self-similarity, high energy protons 
accelerated in the reverse shock may interact with the 10~eV to 1~\keV\ photons 
radiated by the accelerated electrons 
to produce, through pion decay, a burst of duration $\sim t_\pi$ 
of ultra-high energy, $10^{17}-10^{19}$~eV, neutrinos \cite{WnB-AG}
as indicated by Eq.~(\ref{eq:keyrelation}). The flux of these neutrinos
depends on the density of gas surrounding the fireball\index{GRB!circumburst medium}. It is weak, and
undetectable by experiments under construction, if the density is
$n \sim 1$ \cmq, a value typical for the \ISm. If 
GRBs result, however, from the collapse of massive stars\index{GRB!progenitor!collapsar}, then the
fireball is expected to expand into a pre-existing wind and
the transition to self-similar behavior\index{GRB!fireball!self-similar solution} takes place at a radius where the
wind density is $n\approx10^4$ \cmq~ $\gg 1$ \cmq.
In this case, a typical GRB at $z\sim1$
is expected to produce a neutrino fluence \cite{Dai00,WnB-AG}

\begin{equation}
\epsilon_\nu^2\Phi_{\nu_x}\approx
10^{-2.5}\left({\epsilon_\nu\over10^{17}
{\rm eV}}\right)^\alpha~{\rm GeV}~{\rm cm}^{-2},
\label{eq:JAGw}
\end{equation}

\noindent where $\alpha = 0$ for $\epsilon_\nu>10^{17}{\rm eV}$
and $\alpha = 1$ for $\epsilon_\nu<10^{17}{\rm eV}$.
The neutrino flux is expected to be strongly suppressed at energy
$>10^{19}$~eV, since protons are not expected to be
accelerated to energy $\gg10^{20}$~eV. 

\subsection{Implications of Neutrino Emission}
\label{sec:implications}

The predicted intensity of $10^{14}$~eV neutrinos\index{GRB!neutrinos} produced
by photo-meson interactions with observed 1~MeV photons, Eq.~(\ref{eq:JGRB}),  
implies a detection of $\sim10$ neutrino induced muons 
per year in planned $1~{\rm km}^3$ \v{C}erenkov neutrino detectors, 
correlated in time and direction with GRBs \cite{Alvarez00,HnH99,WnB97}. 
The predicted intensity of 
$10^{17}$~eV neutrinos, produced
by photo-meson interactions\index{GRB!fireball!photo-meson} during the onset of fireball interaction 
with its surrounding medium in the case of fireball expansion into 
a pre-existing wind, Eq.~(\ref{eq:JAGw}), 
implies a detection of
several neutrino induced muons per year in a 1 km$^3$ 
detector. In this case, 
the predicted flux of $10^{19}$~eV neutrinos may also be 
detectable by planned large air-shower detectors \cite{Auger-nus,OWL1,OWL2}.

Detection of high energy neutrinos
will test the shock acceleration mechanism and the suggestion that
GRBs are the sources of ultra-high energy protons, since $\ge10^{14}$~eV
($\ge10^{18}$~eV)
neutrino production requires protons of energy $\ge10^{16}$~eV
($\ge10^{19}$~eV). 
The dependence of the $\sim10^{17}$~eV neutrino flux
on fireball environment\index{GRB!circumburst medium} implies that the detection of high energy
neutrinos will also provide constraints on the GRB progenitors. Furthermore,
it has recently been pointed out \cite{Chocked-nus00}
that if GRBs originate from core-collapse of 
massive stars, then a burst of $\ge5$~TeV neutrinos may be produced
by photo-meson interaction while the 
jet\index{GRB!jet} propagates through the envelope, with TeV fluence
implying $0.1-10$ neutrino
events per individual collapse in a $1~{\rm km}^3$ neutrino telescope.
(The neutrino flux which may result
from nuclear collisions in the expanding jet is more difficult to detect
due to the low energy of neutrinos, $\sim10$~GeV, produced by this process 
\cite{BnM00,Derishev99,MnR_nus}).

Detection of neutrinos from GRBs could be used to
test the simultaneity of
neutrino and photon arrival to an accuracy of $\sim1{\rm\ s}$
($\sim1{\rm\ ms}$ for short bursts), checking the assumption of 
special relativity\index{special relativity}
that photons and neutrinos have the same limiting speed.
These observations would also test the weak
equivalence principle, according to which photons and neutrinos should
suffer the same time delay as they pass through a gravitational potential.
With $1{\rm\ s}$ accuracy, a burst at $100{\rm\ Mpc}$ would reveal
a fractional difference in limiting speed 
of $10^{-16}$, and a fractional difference in gravitational time delay 
of order $10^{-6}$ (considering the Galactic potential alone).
Previous applications of these ideas to \SN{1987A}, 
where simultaneity could be checked
only to an accuracy of order several hours, yielded much weaker upper
limits, of order $10^{-8}$ and $10^{-2}$ for fractional differences in the 
limiting speed and time delay respectively \cite{John_book}.

The model discussed above predicts the production of high energy
muon and electron neutrinos. 
However, if the atmospheric neutrino anomaly has the explanation 
usually given \cite{Casper91,Fogli95,Fukuda94},
oscillation to $\nu_\tau$'s with mass $\sim0.1{\rm\ eV}$, then
one should detect equal numbers of $\nu_\mu$'s and $\nu_\tau$'s. 
Up-going $\tau$'s, rather than $\mu$'s, would be a
distinctive signature of such oscillations. 
Since $\nu_\tau$'s are not expected to be produced in the fireball, looking
for $\tau$'s would be an ``appearance experiment.''
To allow flavor change, the difference in squared neutrino masses, 
$\Delta m^2$, should exceed a minimum value
proportional to the ratio of source
distance and neutrino energy \cite{John_book}. A burst at $100{\rm\ Mpc}$ 
producing $10^{14}$ eV neutrinos can test for $\Delta m^2\ge10^{-16}$ eV$^2$, 5 orders of magnitude more sensitive than solar neutrinos.

\subsection*{Acknowledgments} 
This work was supported in part by grants
from the Israel-US BSF (BSF-9800343), MINERVA, and AEC (AEC-38/99).
EW is the Incumbent of the Beracha
foundation career development chair.


\end{document}